\documentclass[12pt]{article}
\usepackage{amssymb, amsmath, amscd}
\usepackage{graphicx}
\usepackage[latin1]{inputenc}
\usepackage{epsfig}
\usepackage{color}
\usepackage{subfig, color}

\newtheorem{definition}{Definition}

\newtheorem{theorem}{Theorem}
\newtheorem{proposition}{Proposition}

\numberwithin{equation}{section}
\newtheorem{example}{Example}
\usepackage{tikz-cd}
\usepackage{rotating}

\begin{document}


\newcommand{\calA}{{\cal A}}
\newcommand{\calB}{{\cal B}}
\newcommand{\calF}{{\cal F}}
\newcommand{\calG}{{\cal G}}
\newcommand{\calR}{{\cal R}}
\newcommand{\calZ}{{\cal Z}}
\def\D{\mathcal{D}}
\def\L{\mathcal{L}}
\def\S{\mathcal{S}}
\def\I{\mathcal{I}}
\def\V{\mathcal{V}}
\def\E{\mathcal{E}}
\def\M{\mathcal{M}}

\def\A{\mathscr{A}}
\def\F{\mathscr{F}}
\def\G{\mathscr{G}}

\newcommand{\N}{{\mathbb N}}
\newcommand{\Z}{{\mathbb Z}}
\newcommand{\Q}{{\mathbb Q}}
\newcommand{\R}{{\mathbb R}}
\newcommand{\CC}{{\mathbb C}}

\newcommand{\K}{{\mathbb K}}
\newcommand{\kk}{{\mathrm k}}

\def\J{\mathrm{J}}
\def\catC{{\bf \mathrm{C}}}
\def\x{\mathrm{x}}
\def\a{\mathrm{a}}
\def\d{\mathrm{d}}
\def\Bi{\mathrm{Bi}}
\def\op{\mathrm{op}}
\def\res{\mathrm{res}}
\def\span{\mathrm{span}}

\newcommand{\C}{{\bf C}}
\newcommand{\Objects}{{\bf Objects}}
\newcommand{\Arrows}{{\bf Arrows}}
\newcommand{\Sets}{{\bf Sets}}

\def\2F1{\mbox{ $_2${F}$_1$}}
\def\1F1{\mbox{ $_1${F}$_1$}}
\def\1F2{\mbox{ $_1${F}$_2$}}
\def\0F1{\mbox{ $_0${F}$_1$}}

\def\GL{\mathrm{GL}}
\def\det{\mathrm{det}}
\def\SL{\mathrm{SL}}
\def\PSL{\mathrm{PSL}}
\def\PGL{\mathrm{PGL}}
\def\O{\mathrm{O}}

\def\gl{\mathfrak{gl}}
\def\g{\mathfrak{g}}
\def\h{\mathfrak{h}}
\def\frakM{\mathfrak{M}}

\newcommand{\Frac}[2]{\displaystyle \frac{#1}{#2}}
\newcommand{\Sum}[2]{\displaystyle{\sum_{#1}^{#2}}}
\newcommand{\Prod}[2]{\displaystyle{\prod_{#1}^{#2}}}
\newcommand{\Int}[2]{\displaystyle{\int_{#1}^{#2}}}
\newcommand{\Lim}[1]{\displaystyle{\lim_{#1}\ }}

\newenvironment{menumerate}{%
    \renewcommand{\theenumi}{\roman{enumi}}%
    \renewcommand{\labelenumi}{\rm(\theenumi)}%
    \begin{enumerate}} {\end{enumerate}}

\newenvironment{system}[1][]%
	{\begin{eqnarray} #1 \left\{ \begin{array}{lll}}%
	{\end{array} \right. \end{eqnarray}}

\newenvironment{meqnarray}%
	{\begin{eqnarray}  \begin{array}{rcl}}%
	{\end{array}  \end{eqnarray}}

\newenvironment{marray}%
	{\\ \begin{tabular}{ll}}
	{\end{tabular}\\}

\newenvironment{program}[1]%
	{\begin{center} \hrulefill \quad {\sf #1} \quad \hrulefill \\[8pt]
		\begin{minipage}{0.90\linewidth}}
	{\end{minipage} \end{center} \hrule \vspace{2pt} \hrule}

\newcommand{\entrylabel}[1]{\mbox{\textsf{#1:}}\hfil}
\newenvironment{entry}
   {\begin{list}{}%
   	{\renewcommand{\makelabel}{\entrylabel}%
   	  \setlength{\labelwidth}{40pt}%
   	  \setlength{\leftmargin}{\labelwidth + \labelsep}%
   	}%
   }%
   {\end{list}}

\newenvironment{remark}{\par \noindent {\bf Remark. }}
			{\hfill $\blacksquare$ \par}

\newenvironment{Pmatrix}
        {$ \left( \!\! \begin{array}{rr} }
        {\end{array} \!\! \right) $}

\newcommand{\fleche}[1]{\stackrel{#1}\longrightarrow}
\def\ssi{si et seulement si\ }
\newcommand{\tab}{\hspace*{\fill}}
\newcommand{\bs}{{\backslash}}
\newcommand{\eps}{{\varepsilon}}
\newcommand{\into}{{\;\rightarrow\;}}
\newcommand{\PD}[2]{\frac{\partial #1}{\partial #2}}
\def\Hat{\widehat}
\def\Bar{\overline}
\def\vect{\vec}
\def\fbar{{\bar f}}
\def\xbar{{\bar \x}}
\newcommand{\afaire}[1]{$$\vdots$$ \begin{center} {\sc #1} \end{center} $$\vdots$$ }
\newcommand{\pref}[1]{(\ref{#1})}

\def\Maple{{\sc Maple}}
\def\RG{{\sc Rosenfeld-Gr\"obner}}



\newcommand{\algf}{\sffamily}
\newcommand{\BEGIN}{{\algf begin}}
\newcommand{\END}{{\algf end}}
\newcommand{\IF}{{\algf if}}
\newcommand{\THEN}{{\algf then}}
\newcommand{\ELSE}{{\algf else}}
\newcommand{\ELIF}{{\algf elif}}
\newcommand{\FI}{{\algf fi}}
\newcommand{\WHILE}{{\algf while}}
\newcommand{\FOR}{{\algf for}}
\newcommand{\DO}{{\algf do}}
\newcommand{\OD}{{\algf od}}
\newcommand{\RETURN}{{\algf return}}
\newcommand{\PROCEDURE}{{\algf procedure}}
\newcommand{\FUNCTION}{{\algf function}}
\newcommand{\INDENTER}{{\algf si} \=\+\kill}

\newcommand{\target}{\mathop{\mathrm{t}}}
\newcommand{\source}{\mathop{\mathrm{s}}}
\newcommand{\trdeg}{\mathop{\mathrm{tr~deg}}}
\newcommand{\jet}[2]{\jmath_{#1}^{#2}}
\newcommand{\rank}{\operatorname{rank}}
\newcommand{\sign}{\operatorname{sign}}
\newcommand{\ord}{\operatorname{ord}}
\newcommand{\aut}{\operatorname{aut}}
\newcommand{\Hom}{\operatorname{Hom}}
\newcommand{\myhom}{\operatorname{hom}}
\newcommand{\codim}{\operatorname{codim}}
\newcommand{\coker}{\operatorname{coker}}
\newcommand{\rp}{\operatorname{rp}}
\newcommand{\leader}{\operatorname{ld}}
\newcommand{\card}{\operatorname{card}}
\newcommand{\Fr}{\operatorname{Frac}}
\newcommand{\RF}{\operatorname{\mathsf{reduced\_form}}}
\newcommand{\rang}{\operatorname{rang}}

\def \Id{\mathrm{Id}}

\def \diff{\mathrm{Diff}^{\mathrm{loc}} }
\def \diffg{\mathrm{Diff} }
\def \Esc{\mathrm{Esc}}

\newcommand{\initial}{\mathop{\mathsf{init}}}
\newcommand{\separant}{\mathop{\mathsf{sep}}}
\newcommand{\quo}{\mathop{\mathsf{quo}}}
\newcommand{\pquo}{\mathop{\mathsf{pquo}}}
\newcommand{\lcoeff}{\mathop{\mathsf{lcoeff}}}
\newcommand{\mvar}{\mathop{\mathsf{mvar}}}

\newcommand{\prem}{\mathop{\mathsf{prem}}}
\newcommand{\remp}{\mathrel{\mathsf{partial\_rem}}}
\newcommand{\remf}{\mathrel{\mathsf{full\_rem}}}
\renewcommand{\gcd}{\mathop{\mathrm{gcd}}}
\newcommand{\pairs}{\mathop{\mathrm{pairs}}}
\newcommand{\dd}{\mathrm{d}}
\newcommand{\ideal}[1]{(#1)}
\newcommand{\cont}{\mathop{\mathrm{cont}}}
\newcommand{\pp}{\mathop{\mathrm{pp}}}
\newcommand{\pgcd}{\mathop{\mathrm{pgcd}}}
\newcommand{\ppmc}{\mathop{\mathrm{ppcm}}}
\newcommand{\init}{\mathop{\mathrm{initial}}}

\bibliographystyle{amsalpha}

{

\title{A Novel Algebraic Geometry Compiling Framework for Adiabatic Quantum Computations}
\author{Raouf Dridi, Hedayat Alghassi, Sridhar Tayur \\ {\small Tepper School of Business,   Carnegie Mellon University,  Pittsburgh, PA 15213}\\
{\small $\{$rdridi, halghassi, stayur$\}$@andrew.cmu.edu}}

\date{\today}
\maketitle

\begin{abstract}
Adiabatic Quantum Computing (AQC) is an attractive paradigm for solving hard integer polynomial optimization problems, as it is robust to environmental noise. Available hardware restricts the Hamiltonians to be of a structure that allows only pairwise interactions, an aspect that will likely remain for the foreseeable future. This requires that the original optimization problem to be first converted -- from its polynomial form~-- to a quadratic unconstrained binary optimization (QUBO) problem, which we frame as a problem in algebraic geometry. Additionally, the hardware graph where such a QUBO-Hamiltonian needs to be embedded  -- assigning variables of the problem to the qubits of the physical optimizer -- is not a complete graph, but rather one with   limited connectivity. This ``problem graph to hardware graph" embedding can also be framed as a problem of computing a Groebner basis of a certain specially constructed polynomial ideal. In this paper, we develop a systematic computational approach to prepare a given polynomial optimization problem for AQC in three steps. The first step reduces an input polynomial optimization problem into a QUBO through the computation of the Groebner basis of a toric ideal generated from the monomials of the input objective function. The second step computes feasible  embeddings. The third step computes the spectral gap of the adiabatic Hamiltonian associated to a given embedding. These steps are applicable well beyond the integer polynomial optimization problem. Our paper provides the first general purpose computational procedure that can be used directly as a {\it translator} to solve polynomial integer optimization. Alternatively, it can be used as a test-bed (with small size problems) to help design efficient heuristic quantum compilers by studying various choices of reductions and embeddings in a systematic and comprehensive manner. An added benefit of our framework is in designing Ising architectures through the study of $\mathcal Y-$minor universal graphs.
\end{abstract}


\medskip 

\newpage
\tableofcontents 

\newpage

\section{Introduction}
 Adiabatic quantum computation (AQC) is a quantum computing paradigm that solves optimization problems of the form 
	\begin{equation}\label{P}
		(\mathcal P):\, argmin_{(y_0, \cdots, y_{m-1})\in \mathbb B ^m} \,  f(y_0, \cdots, y_{m-1}),
	\end{equation}where $\mathbb B = \{0, 1\}$ and $f$  is a polynomial function in $y_0, \cdots, y_{m-1}$  
with rational coefficients~(we write~{$f\in \mathbb Q[y_0, \cdots, y_{m-1}]$}).  In order to do so, each binary variable $y_i$ is mapped into a quantum spin (or a  qubit), and each monomial in $f$  defines 
a many body interaction (or coupling) between the   involved  spins. The collection of these interacting spins defines a quantum system (a self-adjoint $2^m\times 2^m-$matrix) whose energy (spectrum) is exactly the range of the objective function~$f$.  Correspondingly, the solution of the problem $(\mathcal P)$ sits  on the {ground state} (the eigenvector of lowest energy) of 
the quantum system.  AQC finds the ground state  by employing the adiabatic quantum evolution that slowly evolves the ground state of some known system into the sought ground state of the  problem~$(\mathcal{P})$.

~~\\
In reality, the picture is less ideal. Available physical realizations of AQC processors (such as D-Wave Systems processors  \cite{natueDwave})  are built on the {\it Ising model}, where the manufactured qubits are arranged in a three dimensional graph. For instance, Figure \ref{Chimera} depicts
	the arrangement of qubits inside the D-Wave Systems 2000Q processor.  Therein, as in any Ising model, 
	each qubit can  be coupled only with neighboring qubits (2-body interactions), which restricts the function $f$ to a quadratic polynomial, and the problem~$(\mathcal P)$ to a quadratic unconstrained binary optimization (QUBO) problem. Another restrictive feature of current architectures is that
     their hardware graphs are rather non-complete, with limited low edge densities, which makes casting the QUBO into
     a self-adjoint matrix,  as explained above,  highly non-trivial. Theoretically, 
     we know how to overcome both restrictions:  we use   additional variables for the degree restriction and  minor embeddings for 
     the non-completeness restriction. 
     The next examples describe
     this concept of embedding and the difficulties in finding them in practice;  we skip introducing the notion 
     of additional variables  as it is  well known, and for which  have dedicated a section where we relate the important problem of  minimizing their number to toric ideals.   
	 
\begin{figure}[h]
    \centering
    \subfloat  
    {{\includegraphics[width=5cm]{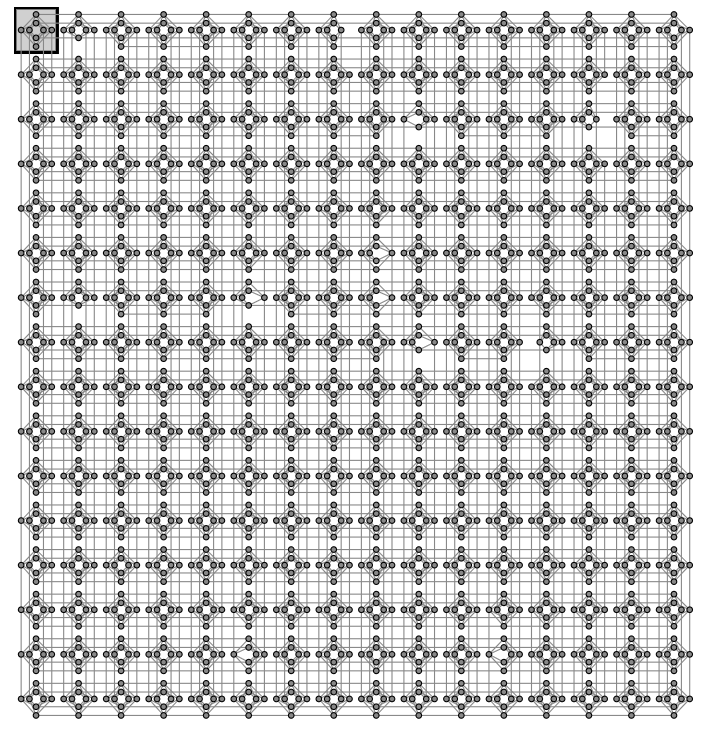} }}%
    \qquad
        \centering
    \subfloat  
    {{\includegraphics[width=8cm]{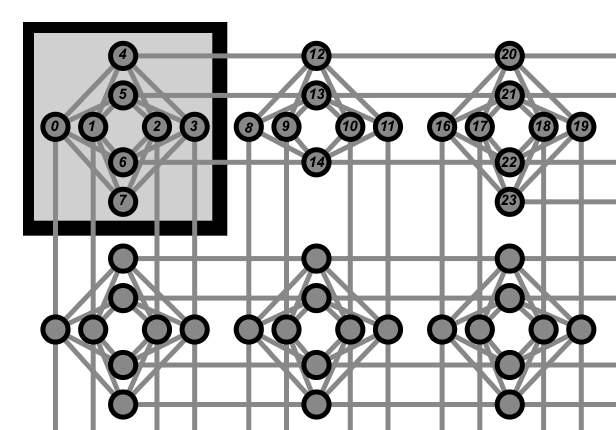} }}%
    \caption{{\small  Cross representation of Chimera $4\times16\times16$, the graph of D-Wave Systems 2000Q processor
     (there is a second representation called column representation  illustrated in the left graph of Figure \ref{chimera2x2Folded}).   Qubits are arranged in $16\times 16$ blocks (or cells). Each block  is a $4\times4$  bipartite graph.
    Qubits have restricted connectivity along the edges where each qubits can interact/couple with at most six neighbors. The missing vertices or edges are faulty qubits or couplers. }}
    \label{Chimera}
  \end{figure}

 \subsection{QUBO to Hardware Embedding: Two illustrative examples}
 
 The first  example serves as an illustration for the notion
 of QUBO to Hardware Embedding; in the second example, we demonstrate our approach and its advantage over current heuristics.  
\subsubsection{First example: What is a QUBO to hardware embedding? }
Consider the following optimization problem that we wish to solve on the D-Wave Systems 2000Q processor: 
	\begin{equation}
		(\mathcal P_\star):\, argmin_{(y_0, \cdots, y_{m-1})\in \mathbb B ^m} \quad  y_0 \sum_{i=1}^8 c_i y_i. 
	\end{equation}
	 \begin{figure}[h]
    	\centering
    		\subfloat  
    		{{\includegraphics[width=3cm]{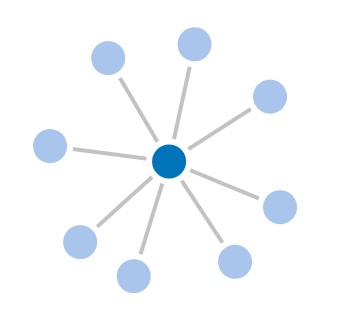} }}%
    		\caption{{\small The logical graph of the objective function in $(\mathcal P_\star)$ }}
  	  	\label{star}
 	 \end{figure}%
As mentioned before, this requires assigning the problem variables $y_0,\cdots, y_8$ to the physical qubits of the processor.      
Let us denote by $X$   the {\it physical} graph depicted in Figure~\ref{Chimera} and 
by $Y$    the {\it logical} graph that represents the quadratic  objective function of the problem~$(\mathcal P_\star)$ given in Figure~\ref{star}.  We can start  
by attempting to  embed  $Y$ inside $X$  by  matching edges to edges; we quickly realize that this is not an option, because  the degree of the vertex $y_0$ is 8 whilst  the maximum degree  in Chimera graph $X$ is 6.  On the other hand,   by drawing an analogy to
the  blowup procedure in algebraic geometry, we can blowup the singularity -- here  the intersection point $y_0$ -- into a line (a chain).
Figure \ref{blowup} depicts three  {\it minor embeddings} of $Y$, where the problem qubit
       $y_0$ is represented by a chain of physical qubits.    In fact, a minor embedding can be understood as  a sequence of blow ups of high degree or distant vertices. For more complicated problems, finding the right sequence of blow ups {(assuming one exists)} is non-trivial and is a central question of this paper.
       The next example discusses this point. 
 \begin{figure}[h]
    \centering
    \subfloat  
    {{\includegraphics[width=4cm]{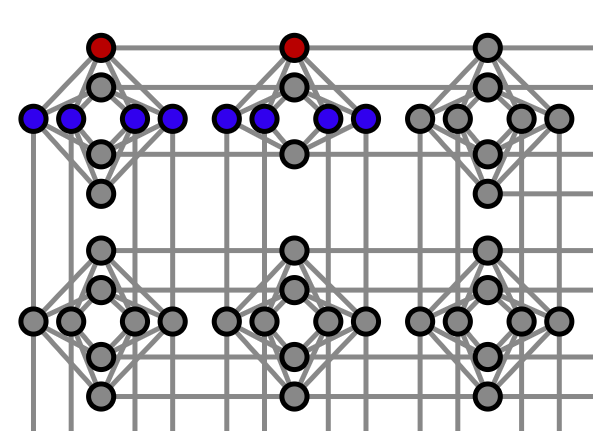} }}%
        \centering
    \subfloat  
    {{\includegraphics[width=4cm]{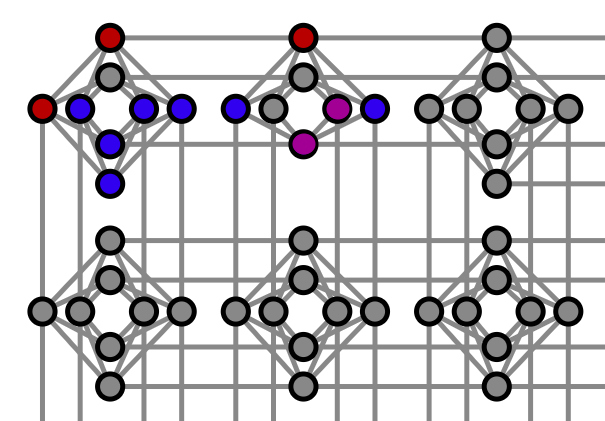} }}%
        \centering
    \subfloat  
    {{\includegraphics[width=4cm]{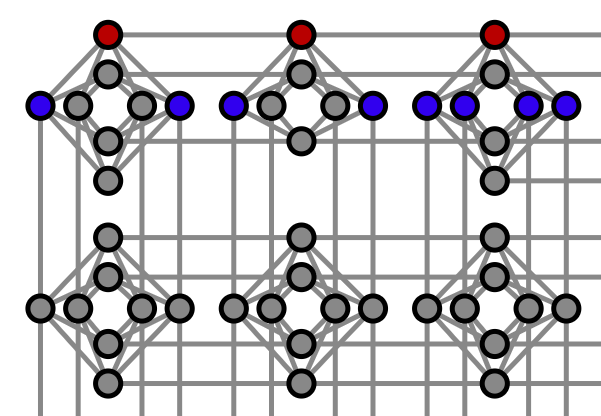} }}%
    \caption{{\small Three different examples of embeddings of the logical graph \ref{star} inside Chimera.  In all cases, the red chains of qubits represent the logical qubit $y_0$. The remaining qubits
    are represented with 1-chain (i.e., a physical qubit) in blue, except for the middle picture, where $y_7$ is represented by the purple 2-chain.  }}
    \label{blowup}
  \end{figure}

 \subsubsection{Second example: Demonstration of our approach}
  In this example, the logical graph  $Y$  is given in Figure \ref{exple2Intro},  and we wish, as before, to embed it
   inside the Chimera graph.  The embedding is not hard to find by inspection: the right block
   of the graph $Y$ is a $4\times 4$ bipartite cell in Chimera (left graph of Figure \ref{chimera2x2Folded}), and we can embed the remaining left block (the two  triangles)
   by collapsing  (at least) one edge of a second   neighboring cell.  The subtlety here, however, is that the only way to embed the graph $Y$ inside a 2-blocks Chimera is
   by collapsing  edges that are entirely contained inside one of the blocks. Any heuristic that looks
   for these  chains otherwise will fail -- it is easy to scale the example, making it very hard for current heuristics. 
   
     \begin{figure}[h]
    \centering
    \subfloat  
    {{\includegraphics[width=5cm]{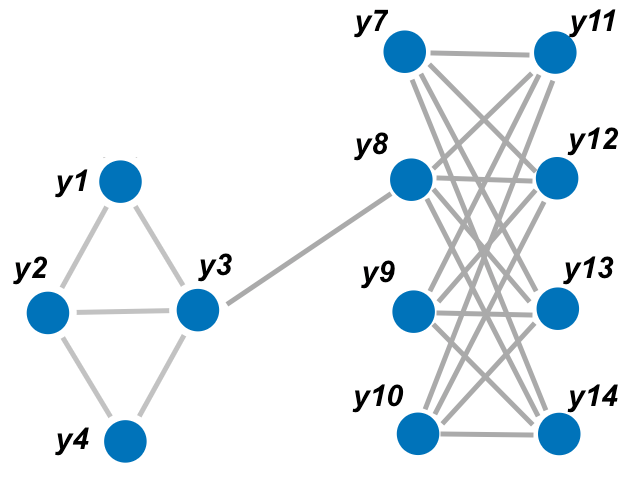} }}%
    \caption{{\small To embed the depicted problem graph inside Chimera $2\times 2$ (left graph of Figure \ref{chimera2x2Folded}),  we need to collapse edges that are entirely contained inside one of the blocks.  Heuristics that looks
   for these  chains otherwise will fail.   }}
    \label{exple2Intro}
  \end{figure} 

~~\\ 
We  can obtain this embedding  as follows. First, we think of an embedding as a surjection~$\pi$ from the hardware graph $X$ to the logical graph $Y$ -- the 
triplet $(X, Y, \pi)$ forms a {\it fiber-bundle}.  This surjection $\pi$ maps a chain of physical qubits (a {\it fiber}) into at most one logical qubit. Or,   if we   write
\begin{equation}
	\pi(x_i) = \sum_{y_j \in {\bf Vertices}(Y)} \alpha_{ij} y_j
\end{equation}
for each   $x_i \in {\bf Vertices}(X)$,  then at most one of  the binary numbers $\alpha_{ij}$  is 1 for each $x_i $ (if the qubits $x_i $ is not used then we set $\pi(x_i)=0$). So, the condition
that chains should not intersect (that is, $\pi$ is a well defined map) translates into a set of algebraic equations on the parameters $\alpha_{ij}$.   
Similarly,  all the requirements that an embedding needs to satisfy  (which we will review in Subsection \ref{Embeddings}) can be formulated  as a set of algebraic equations on the binary parameters $\alpha_{ij}$ (presented in Subsection \ref{FiberBundles}).   Therefore, the set of {\it all} embeddings  (up to any desired size) is given by the set of zeros (algebraic variety) of this system.   When this   variety  is empty,  the logical graph $Y$ is not embeddable inside~$X$.  

~~\\
In our example of Figure \ref{exple2Intro},  the variety is not empty and it  suffices to solve its defining system to obtain the sought embedding of the graph $Y$. However, we can do much better that this  by employing invariant theory to compress the different algebraic expressions -- many of the solutions are redundant  (identical up to symmetries in~${\bf Aut}(X)$), 
which affects the efficiency of the method for large graphs. Indeed, we can get rid of this redundancy by {\it folding} the hardware graph $X$ along its symmetry axis as in Figure~\ref{chimera2x2Folded} -- this folding operation is made precise and systematic in  \ref{symmetries}, where we re-express its quadratic form in terms of the invariants of the symmetry. The quadratic form of the new graph depicted in  Figure \ref{chimera2x2Folded} is
\begin{equation} 
            K_1   K_3 + K_1   K_4 + K_2   K_3 +   K_2   K_4 +    K_4   K_5 +  K_3   K_6 +  K_8   K_5 + 
	  +  K_8   K_6 +  K_7   K_5 +  K_7   K_6. 
\end{equation}
The nodes $K_i$ are the invariants  of the symmetry (see also the caption of Figure~\ref{chimera2x2Folded})
 \begin{eqnarray}
 	K_1 &=& x_{1}+x_{2},\quad \, K_2 = x_{3}+x_{4},\quad \,\, \, K_3 = x_{7}+x_{8},\quad K_4 = x_{5}+x_{6},\\
	K_5 &=& x_{9}+x_{10},\quad K_6 = x_{11}+x_{12},\quad K_7 = x_{15}+x_{16},\quad K_8 = x_{13}+x_{14}.
 \end{eqnarray}
  The map $\pi$ now takes the form
 \begin{equation}
 	\pi(K_i) = \sum_{y_j \in {\bf Vertices}(Y)} \alpha_{ij} y_j,
 \end{equation}
where the goal  is to embed  the problem graph of Figure \ref{exple2Intro} inside this folded Chimera.  
 We generate the system of equations on the parameters $\alpha_{ij}$ with this new target graph and  solve. We  obtain the solution
 \begin{system}
 	K_1 &=& y_1+y_2, \\
	K_2 &=& K_3= y_4,\\
	K_4 &=& y_1 + y_3,  \\
	K_5 &=&  y_7 +y_8, \\
	K_6 &=& y_9+y_{10},\\
	K_7 &=& y_{13}+y_{14},\\
	K_8 &=& y_{11}+y_{12}. 
\end{system}%
The first equation says that the formal sum $x_{1}+x_{2}$ maps to the formal sum  $y_1+y_2$. Any choice of assigning values to $x_1$ and $x_2$
is equally valid -- they were redundant before the use of invariants.  The remaining equations are treated similarly. The second equation says that the nodes $K_2=x_{3}+x_{4}$ and $K_3=x_{7}+x_{8}$ collapse into the   qubits $y_4$ hence, for instance,  the edge $(x_3, x_7)$ collapses into $y_4$. Collapsing any of the edges $(x_3, x_8)$,  $(x_4, x_7)$ and  $(x_4, x_8)$ is also valid but redundant. 


    \begin{figure}[h]
    \centering
    \subfloat  
    {{\includegraphics[width=4cm]{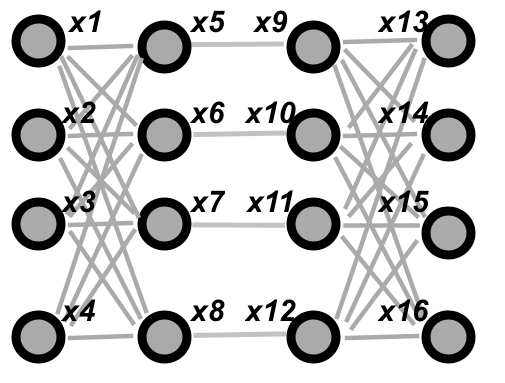} }}%
        \centering
    \subfloat  
    {{\includegraphics[width=7cm]{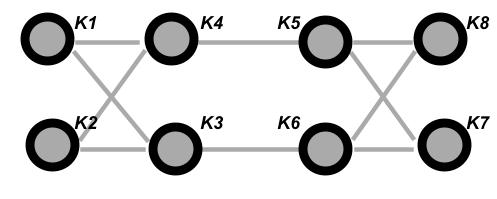} }}%
    \caption{{\small  The left graph is (the column representation of) Chimera $2\times 2$. The permutation that exchanges
    the chain $[x_1, x_5, x_9, x_{13}]$ with $[x_2, x_6, x_{10}, x_{14}]$ and the chain
    $[x_3, x_7, x_{11}, x_{15}]$ with $[x_4, x_8, x_{12}, x_{16}]$,  
     is a symmetry.  The right graph is obtained by re-expressing the quadratic form of the left graph in terms of the invariants of this symmetry. Embedding the problem graph in Figure~\ref{exple2Intro} inside this folded version of Chimera  leads to more efficient calculations.}}
    \label{chimera2x2Folded}
  \end{figure}
  
\subsubsection{More intricacies of finding embeddings in the context of AQC}
The two examples above illustrate  how difficult is the problem of embedding from the algorithmic point of view. What makes this problem even more difficult is the fact that not all  minor embeddings are equally useful for AQC. First,  the number of  physical qubits used is  important  (recall
  that the dimension of the Hilbert space is  exponential in the number of qubits). Second, the size of the chains -- the number of replications of an individual problem qubit that need to be linked together to form a chain -- as well as their couplings
  has significant implications on the effectiveness of the embedding (See Figure \ref{chainBreaking}). Third, 
  as we show in this paper, the theoretical computational speedup in AQC itself depends on the choice of the minor embedding.  
  
   \begin{figure}[h]
    \centering
    \subfloat  
    {{\includegraphics[width=6cm]{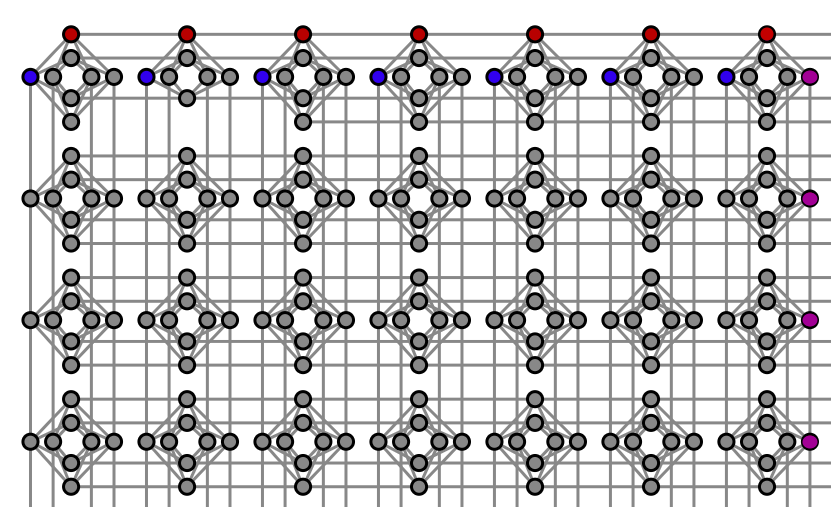} }}%
    \caption{{\small The depicted minor embedding (for  the problem $(\mathcal P)_*$) has two long chains that don't persist through the adiabatic evolution (in D-Wave). In this case, the quantum processor fails to return an answer.   }}
    \label{chainBreaking}
  \end{figure}
  

\subsection{Goals of the paper}
Suppose we are given a polynomial binary optimization problem
\begin{equation}
	(\mathcal P):\, argmin_{(y_0, \cdots, y_{m-1})\in \mathbb B ^m} \,  f(y_0, \cdots, y_{m-1}).
\end{equation}
In this paper, using a novel  algebraic geometry perspective on AQC, we make three contributions: 
\begin{itemize}
	\item [{\bf 1}]  An end-to-end systematic procedure for solving optimization problems $(\mathcal P)$, using Ising AQC processors. This {\em translator} (or compiler) can be programmed classically starting from the problem model and ending up with an input into the quantum computer.  One can view this as the first fully systematic compiler for AQC. 
    \item [{\bf 2}] Computing the spectral gap of the adiabatic Hamiltonian as an algebraic function of the points of the embedding variety.
	\item[{\bf 3}] A systematic procedure for the design of near term Ising architectures.  
	This design problem is referred to as the $\mathcal Y-$minor universal problem in the literature  \cite{Choi2011}. The task is to design architectures that obey the physical engineering constraints (low connectivity of the manufactured qubits) and still are able  to tackle interesting, hard problems.  
	\end{itemize}
We recognize that the (worst-case) computational complexity of our procedure is not polynomial. Indeed, at this time, we are interested  primarily in developing a robust theoretical framework that can form the basis of a computational procedure that can be programmed into software, and laying out the various issues that arise as we move from optimizing polynomials on lattices to creating embeddable Hamiltonians on physically realizable architectures. This allows us to study small problem instances in a systematic and comprehensive manner on actual physical devices. Thus, our framework can serve as a sandbox to test   various heuristics to help design an efficient and scalable (that may have a provable worst-case polynomial time performance) quantum compiler. Similarly, through the study of the $\mathcal Y-$minor universal problem, we can help design good, physically realizable hardware.
  

\begin{center}
\begin{itemize}
	\item[] {\bf Flowchart of the Translator: }
	\item[$\rightarrow$] {\sf The user inputs the optimization problem} $(\mathcal P)$. 
	\item[A] {\sf Reduction to a quadratic form:} (Details in Section 4.1)
	  \begin{itemize}
	  \item[1] {\sf Generation of the toric ideal $\mathcal J_A$ from the monomials of the objective function of $(\mathcal P)$.} 
	  \item[2] {\sf Computation of a reduced  Groebner basis for $\mathcal J_A$; return the quadratic 
	  function~${\mathcal E \in \mathbb Q[y]}$.}
	\end{itemize}
	
	\item[~~ B] {\sf Embedding inside the  AQC processor graph:} (Details in Section 4.2)
	  \begin{itemize}
\item[3]  
	{\sf Generation of the ideal $\mathcal I$ that gives the minor embeddings $\pi$}. 
	\item[4] {\sf Computation of a  reduced  Groebner basis $\mathcal B$ of the ideal $\mathcal I$}. 
	\item[5] {\sf  If $1\in \mathcal B $. Go back to 2 and choose a different quadratic function.}
	\item[6] {\sf Comparison of the different embeddings with respect to their effect on the spectral gap}. (Details in Section 4.3)
	\end{itemize} 
	\item[~ ~ C] {\sf Solution using a selected embedding on the AQC processor}. 
	\item[$\leftarrow$] {\sf User gets the answer.} 
\end{itemize}
\end{center}

\subsection{Outline of the Paper}

The paper is structured as follows.  Section 2 briefly summarizes AQC on Ising spin glass architectures. Section 3 briefly reviews Groebner bases, toric ideals, and related useful results. Section 4.1 uses toric ideals to provide an algorithm for the reduction step, resulting in a QUBO.  Section 4.2 
details  the calculation of minor-embedding using Groebner bases. We show that the set of all embeddings  is an algebraic variety.  The automorphisms  of the hardware graph, specifically their  invariants,  are used to help
with the computation. We also show how the number of minor-embeddings is determined using staircase diagrams. Section 4.3 provides a method to compute the spectral gap as an algebraic function of the points of the embedding variety. In section 5, we solve the problem of $\mathcal Y$-minor embedding universal \cite{choi} using Groebner bases.  We conclude in Section 6.


~~\\
{\bf Notations} 

~~\\
All graphs considered here are simple and undirected. 
The following notation is used in the remainder of the paper: 
\begin{itemize}
	\item ${\bf Vertices}(X)$ and ${\bf Edges}(X)$ are the vertex and edge sets of the graph $X$. 
	\item $n = card({\bf Vertices}(X))$ is the size of the hardware graph $X$.
	\item $m = card({\bf Vertices}(Y))$ is the size of  the problem graph $Y$.
	\item ${\bf Vertices}(X) = \{x_0,\cdots, x_{n-1}\}$ and $x=(x_0,\cdots, x_{n-1})$.
	\item ${\bf Vertices}(Y) = \{y_0,\cdots, y_{m-1}\}$ and $y=(y_0,\cdots, y_{m-1})$. 	
	\item $\mathbb Q[x_0,\cdots, x_{n-1}]$ the ring of polynomials in $x_0,\cdots, x_{n-1}$ with rational coefficients. 
	\item $Q_X(x) := \sum_{x_{i_1}x_{i_2} \in {\bf Edges}(X)} x_{i_1}x_{i_2} \in \mathbb Q[x_0,\cdots, x_{n-1}] $ is the quadratic form of the graph~$X$.
	\item $Q_Y(y) := \sum_{y_{j_1}y_{j_2} \in {\bf Edges}(Y)} y_{j_1}y_{j_2} \in \mathbb Q[y_0,\cdots, y_{m-1}] $  is the quadratic form of the graph~$Y$.
	\item $\alpha = (\alpha_{ij}, \, i=0..n-1, j=0..m-1)$.
	\item $\beta = (\beta_0, \cdots, \beta_{n-1})$.
	\item $u=(u_0, \cdots, u_{n-1})$ and $x^u = x_0^{u_0}\cdots  x_{n-1} ^{u_{n-1}}$. 
\end{itemize}

\section{The Physics: AQC on Ising Spin Glass Architectures} 
The primary purpose of Adiabatic quantum computation (AQC) \cite{Farhi472, Nishimori} is to 
 solve the  problem of computing  ground states   of high dimensional Hamiltonians  (self-adjoint operators acting on large Hilbert spaces, usually~${{\mathbb C^2}^{\otimes n}:= {\mathbb C^2}\otimes\cdots\otimes {\mathbb C^2}}$).
It is a straightforward application of the adiabatic theorem \cite{Born1928, Kato, Avron1999} to time dependent 
Hamiltonians of the form 
\begin{equation}
H(t) =  (1-t/T) H_{initial} + tH_{(\mathcal P)}.
\end{equation}
  This  theorem  states that the quantum system, initialized at the known ground state of the initial Hamiltonian $H_{initial}$,  will  remain at the ground state of  $H(t)$ for all time $t\leq T$ with probability inversely proportional to the square of the energy difference with the rest of the spectrum.     A simple series expansion  of the wave function in the slow regime (see for instance \cite{Nishimori})
   shows that the total time $T$ spent to  adiabatically attain the sought ground state is 
  \begin{equation}
 	T \sim O\left( \frac{1}{min_{0\leq t\leq T}  \Delta(t)} \right)
 \end{equation}
 where $\Delta(t)$ is the gap between the two smallest eigenvalues (spectral gap) of $H(t)$.  A measurement of the final state will yield a solution of the problem.  A notable fact about AQC  is that 
 it enjoys 
proved robustness   against environment noise \cite{PhysRevA.65.012322, PhysRevA.74.052322} (as long as the temperature of the environment is not too high),  making AQC a reasonable choice for near term quantum computing, 

~~\\ 
As mentioned before, only  the restricted class of 
 Ising spin glass Hamiltonians (Ising Hamiltonians for short)  is currently physically realized. The quantum system is constituted of a set of
spins  that can point to two directions and are  arranged in a graph $X$ where only  local 2-body interactions, along the edges of $X$,  are allowed (See Figure \ref{Chimera}). More formally, Ising Hamiltonians  are of the form 
\begin{equation}
	H_{(\mathcal P)} = \sum_{i\in {\bf Vertices}(X)} h_i \sigma_i^3 + \sum_{ij\in {\bf Edges}(X)} J_{ij} \sigma^3_i \sigma^3_j,
\end{equation}
with   $\sigma_i^3=I\otimes\cdots\otimes \sigma^3\otimes \cdots \otimes I$  
and $\sigma^3_i \sigma^3_j=I\otimes\cdots\otimes \sigma^3\otimes \cdots\otimes \sigma^3\otimes \cdots \otimes I$. Here,  $\sigma^3$ is the Pauli operator 
\begin{equation}
	\sigma^3 = 
	\begin{bmatrix}
    1      & 0 \\ 
    0    & -1
\end{bmatrix}
\end{equation}
and $I$   is the $2\times 2$ identity matrix.  The
  coefficients $h_i$ are the  biases and
the coefficients $J_{ij}$, called couplings,  determine the strength of the interactions  
between the two spins $\sigma^3_i $ and $\sigma^3_j$. When the coupling  $J_{ij}$ is negative, in which case we say $J_{ij}$ is a ferromagnetic coupling, the two spins tend to point to  the same direction. Inversely, when $J_{ij}$
is positive,  anti-ferromagnetic coupling, the two spins  point to opposite directions. 
With different coupling strengths and signs, the aggregated interaction is, in general, complicated, and computing  the ground state of 
$H_{(\mathcal P)} $ is
  NP \cite{0305-4470-15-10-028, 10.3389/fphy.2014.00005}. Note that, with the restriction to Ising architecture,  AQC is  no longer universal \cite{1366223,PhysRevLett.99.070502}.

~~\\    
 The Hamiltonian $H_{(\mathcal P)}$ is intentionally made diagonal in 
  the computational basis\footnote{
  This basis is given by the eigenstates of
  the Pauli operators $\sigma_i^3$. Explicitly, vectors of the computational basis are states $ | y_1 \cdots y_m\rangle =  | y_1\rangle\otimes \cdots \otimes|y_m\rangle \in {\mathbb C^2}^{\otimes n}$
 where $|0\rangle = (1, 0)\in \mathbb C^2$ and $|1\rangle = (0,1)\in \mathbb C^2$  
  are the two eigenstates  of $\sigma^3$
with eigenvalues 1 and -1, respectively.}; that is, 
vectors of the computational basis are eigenstates of
$H_{(\mathcal P)}$.  Consequently, 
calculating
the scalar products $\langle y_1 \cdots y_m|H_{(\mathcal P)}| y_1 \cdots y_m\rangle$ gives the energy function of $H_{(\mathcal P)}$: 
\begin{equation}
	\mathcal E_{(\mathcal P)} (s_1, \cdots, s_n) = \sum_{i\in {\bf Vertices}(X)} h_i s_i + \sum_{ij\in {\bf Edges}(X)} J_{ij} s_{i_1} s_{i_2},
\end{equation}
where    $s_i \in \{-1, 1\}$ or by taking $x_i = (s_i +1)/2$  (consistently with 01-notation in binary optimization): 
\begin{equation}\label{energyHardware}
	\mathcal E_{(\mathcal P)} (x_1, \cdots, x_n) = \sum_{i\in {\bf Vertices}(X)}  h_i x_i + \sum_{ij\in {\bf Edges}(X)} J_{ij} x_{i_1} x_{i_2},
\end{equation}
where we have used the same notations $h_i$ and  $J_{ij}$ for the new adjusted coefficients.  The energy function measures the violations, by the given spin configuration, of the ferromagnetic and anti-ferromagnetic couplings. 
The ground state of the Hamiltonian $H_{(\mathcal P)}$ coincides with the spin configuration with the minimum amount of violations, which is also the global minimum of the energy function.



\begin{table}[h]
\centering
\begin{tabular}{|c|c|}
\hline\\
Optimization problems & Adiabatic quantum computations\\
\hline \hline\\
Polynomial binary optimization & Many-body (many-spin) Hamiltonian\\
\hline\\
Quadratic binary optimization &  Ising Hamiltonian\\
\hline\\
Objective function & Energy function\\\hline\\
Binary variables  $y_i$&  Qubit $|y_i\rangle$ i.e., state of the $i$th spin $\sigma^3$ \\\hline\\ 
Monomials  $J_{ij}y_iy_j$ & Coupled spins with Coupling strengths $J_{ij}$\\\hline\\
Global minima & Ground state (possibly degenerate)\\\hline\\
Local minima & States (spectrum) of the Hamiltonian \\\hline\\
Search space $\{0, 1\}^n$& Hilbert space ${\mathbb C^2}^{\otimes n}$\\
\hline
\end{tabular}
\caption{Correspondence between optimization problems and adiabatic quantum computations.}
\end{table}
 
~~\\
There are several proposals for the initial Ising Hamiltonian $ H_{initial}$,  all with the property that  cooling to their ground states is easy.  Our exposition doesn't depend on the choice of the Hamiltonian~$H_{initial}$.


\section{The Mathematics: Groebner Basis and Toric Ideals} 
The intertwining between algebraic geometry and optimization is a fertile research area.  The collective work
of  B. Sturmfels and collaborators \cite{MR1363949, realalggeo}  is of particular interest. The application of algebraic geometry to integer programming can be found  in \cite{Conti:1991:BAI:646027.676734, DBLP:journals/mp/TayurTN95,  DBLP:journals/mp/SturmfelsT97,  doi:10.1287/mnsc.46.7.999.12033}.  Sampling from conditional distributions is suggested in \cite{diaconis1998}.  Application to prime factoring in conjunction with AQC is explored in~\cite{raouffactorization}. 

~~\\ 
Let $\mathcal S$ be 
 a  set of polynomials $f\in \mathbb Q[x_0, \ldots, x_{n-1}]$. Let  $\mathcal V(S)$ denotes the affine algebraic variety defined by 
the polynomials $f\in S$, that is, the set of common zeros of the equations $f=0, \, f\in \mathcal S$. The system $\mathcal S$ 
generates an ideal $\mathcal I $ by  taking all linear combinations over $\mathbb Q[x_0, \ldots, x_{n-1}]$ of all polynomials in  $\mathcal S$; we have $\mathcal V(\mathcal S)=\mathcal V(\mathcal I).$ The ideal $\mathcal I$ reveals the hidden polynomials that are the consequence of the generating  polynomials in $\mathcal S$. For instance, if one of the hidden polynomials is the constant polynomial 1 (i.e., $1\in \mathcal I$), then the system $\mathcal S$ is inconsistent (because $1\neq 0$).

~~\\ 
Strictly speaking, the set of all hidden polynomials is given
by the so-called radical ideal $\sqrt{\mathcal I}$, which is defined by \mbox{$\sqrt{\mathcal I} = \{g \in\mathbb Q[x_0, \ldots, x_{n-1}] |\,  \exists r\in \mathbb N: \, g^r\in \mathcal I \}$}.  In practice,  the ideal $\sqrt{\mathcal I}$ is infinite, so we represent such an ideal
 using a Groebner basis $\mathcal B$, which one might take to be a triangularization of the ideal $\sqrt{\mathcal I}$.  In fact, the computation of   Groebner bases  generalizes Gaussian elimination in linear systems. We also have $\mathcal V(\mathcal S)=\mathcal V(\mathcal I)=\mathcal V(\sqrt{\mathcal I})=\mathcal V(\mathcal B)$  and (Hilbert's Nullstellensatz theorem:) $\mathcal I(\mathcal V(\mathcal I))=\sqrt{\mathcal I}$.

~~\\
 {\it Term orders.} A term order  on $\mathbb Q[x_0, \ldots, x_{n-1}]$ is a total order $\prec$ on the set of all
monomials $x^a=x_1^{a_1}\ldots x_n^{a_n}$, which has the following properties:
\begin{itemize}
\item    if $x^a\prec x^b$, then $x^{a+c}\prec x^{b+c}$ for all positive integers $a, b$, and $c$;  
\item   $1\prec x^a$ for all strictly positive integers $a$.
\end{itemize}
 An example of this is the pure lexicographic order plex $x_0 \succ x_1 \succ \cdots$. Monomials
are compared first by their degree in $x_0$, with ties broken by degree in $x_1$, etc. This order is usually used in eliminating variables.  Another example, is
the graded reverse lexicographic order tdeg. Monomials are compared first by their total degree, with ties broken by reverse lexicographic order. This order typically provides faster Groebner  basis computations.
 
~~\\
{\it Groebner bases.}
Given a term order $\prec$ on $\mathbb Q[x_0, \ldots, x_{n-1}]$, then by the leading term (initial term) {\sf LT} of $f$ we mean the largest monomial in $f$ with respect to $\prec$. A reduced Groebner basis to the ideal $\mathcal I$ with respect to the ordering~$\prec$ is a subset $\mathcal B$ of $\mathcal I$ such that: 
\begin{itemize}
\item   the initial terms of elements of $\mathcal B$ generate the ideal ${\sf LT}(\mathcal I)$ of all initial terms of~$\mathcal I$; 
\item   for each $g\in \mathcal B$, the coefficient of the initial term of $g$ is 1; 
\item   the set ${\sf LT}(g)$ minimally generates ${\sf LT}(\mathcal I)$; and
\item   no trailing term of any $g\in \mathcal B$ lies in ${\sf LT}(\mathcal I)$.
\end{itemize}
 Currently, Groebner bases are computed using sophisticated versions of the original Buchberger algorithm, for example, the F4  and F5 algorithms  by J. C. Faug\`ere \cite{Faugere199961}.

\begin{theorem}
Let $\mathcal{I}\subset \mathbb Q[x_0, \ldots, x_{n-1}]$ be an ideal and let $\mathcal{B}$ be a reduced Groebnber basis of~$\mathcal{I}$
with respect to the lex order $x_0\succ \ldots \succ x_{n-1}$. Then, for every $0\leq l\leq n-1$, the set 
\begin{equation}\label{intersectionB}
	\mathcal{B}\cap \mathbb Q[x_{l}, \ldots, x_{n-1}]
\end{equation}
is a reduced Groebner basis of the ideal $\mathcal{I}\cap Q[x_{l}, \ldots, x_{n-1}]$.
\end{theorem}
We shall use this elimination theorem  repeatedly in this paper. It is used to obtain
the complete set of conditions on the variables $x_{l}, \ldots, x_{n-1}$ such that the ideal $\mathcal{I}$ is not empty. For instance, if the ideal represents a system of algebraic equations and these equations are (algebraically) dependent on certain parameters, then the  intersection \pref{intersectionB}
gives {\it all} necessary and sufficient conditions for the existence of solutions. 

~~\\
 {\it Normal forms.} A normal form is the remainder of Euclidean divisions in the ring of polynomials $k[x_0, \ldots, x_{n-1}]$. 
 Precisely, let $\mathcal B$ be a reduced  Groebner basis for an ideal $\mathcal I\subset k [x_0, \ldots, x_{n-1}]$. 
The normal form of a polynomial $f\in \mathbb Q[x_0, \ldots, x_{n-1}]$, with respect to  
$\mathcal B$,  is the unique   polynomial~${{\sf NF}_\mathcal B(f)\in \mathbb Q[x_0, \ldots, x_{n-1}]}$ that satisfies the following
properties: 
\begin{itemize}
\item  No term of ${\sf NF}_\mathcal B(f)$ is divisible by an ${\sf LT}(p), \, p\in \mathcal B$ 
\item   There is $g\in \mathcal I$
such that $f=g+ {\sf NF}_\mathcal B(f)$. Additionally, ${\sf NF}_\mathcal B(f)$  is the remainder of division of $f$ by $\mathcal B$
no matter how elements of $\mathcal B$ are listed when performing the Euclidean division.
\end{itemize}
The remainder ${\sf NF}_\mathcal B(f)$ is the
canonical  representative for the equivalence class of $f$ modulo~$\mathcal I$. 
If one knows that $1\notin \mathcal I$, the generalized division algorithm ~(\cite{Cox:2007:IVA:1204670}, Chapter~2) can be applied directly using $\mathcal I$ and the given monomial order, without computing Groebner bases. However, in this case,   the result is not always equal to the canonical remainder  
  that $\mathcal B$ gives.

~~\\ 
{\it Toric ideals.} These are ideals
 generated by differences of monomials. Their Groebner bases enjoy a clear structure  given by   kernels of integer matrices.  Specifically,  let $A =(a_1, \cdots, a_n)$ be any integer $m\times n$-matrix ($A$ is called configuration matrix). Each 
 column $ {{\bf a}}_i = (a_{1i}, \cdots, a_{ni})^T$ is identified with  a Laurent monomial $y^{ {{\bf a}}_i} = y_1^{a_{1i}}\cdots y_m^{a_{ni}}$.
 The toric ideal $\mathcal J_A$ associated   with the configuration $A$ is the kernel of the algebra homomorphism 
 \begin{eqnarray}
	\mathbb Q[x] \rightarrow \mathbb Q[y]\\
	 x_i \mapsto y^{ {{\bf a}}_i}.
 \end{eqnarray}
 We have: 
 \begin{proposition}\label{ker}
 The toric ideal  $\mathcal J_A$ is generated by the binomials $x^{{{\bf u}}_+} - x^{{{\bf u}}_-},$ where the vector 
  ${\bf u} = {{\bf u}}_+ - {{\bf u}}_{-}\in {\mathbb Z^+}^n\oplus {\mathbb Z^+}^n$
  runs over all integer vectors in  $\mathrm{Ker}_\mathbb Z A$, the  kernel of the matrix~$A$.	
 \end{proposition}
  Assuming $A\subset \mathbb N^n$, a conceptually easy method for computing generators of $\mathcal J_A$ is to use 
  the elimination theorem as follows (more efficient algorithms, based on Proposition \ref{ker}, are described in \cite{MR1363949}):  Consider the polynomial ring
  $\mathbb Q[x, y]$ and define the ideal $\mathcal K_A$ of   $\mathbb Q[x, y]$ by
  \begin{equation}
  	\mathcal K_A = < x_1 - y^{{{\bf a}}_2}, \, x_2 - y^{{{\bf a}}_2}, \, \cdots, x_n - y^{{{\bf a}}_n}>. 
  \end{equation}
  The toric ideal $\mathcal J_A \subset \mathbb Q[x]$ of $A$ is equal to the intersection of the ideal $\mathcal K_A\subset \mathbb Q[x, y]$ and the ring $\mathbb Q[x]$; that is $\mathcal J_A = \mathcal K_A\cap \mathbb Q[x].$  If we consider the plex order on   
  $\mathbb Q[x, y]$ 
  induced by the ordering $y_i \succ x_i$ and compute the reduced Groebner basis $\mathcal B$ of $\mathcal K_A$ with respect to this plex,
  then  
  the intersection $\mathcal B \cap  \mathbb Q[x]$ is the reduced Groebner basis of $\mathcal J_A$. In particular, $\mathcal B \cap  \mathbb Q[x]$   is a system of generators of~$\mathcal J_A$.

\section {The Mathematics that Enables Physics to Optimize Polynomial Programs}  
In the context of AQC, compiling the binary optimization problem 
 \begin{equation}
	(\mathcal P):\, argmin_{(y_0, \cdots, y_{m-1})\in \mathbb B ^m} \,  f(y_0, \cdots, y_{m-1})
\end{equation}
into the hardware graph consists of two steps: 1)~reducing the objective function of $(\mathcal P)$  into
quadratic function and 2) embedding this function into the hardware graph. 
In this long section, we algorithmize  these two steps  using   algebraic geometry. We start with the first: 

\subsection{Reductions to quadratic optimization and toric ideals} \label{reduction}
Consider the step of reducing the polynomial optimization~$(\mathcal P)$
to a quadratic optimization.  
We have explained in the introduction that this is a key step in the process of  mapping the problem~$(\mathcal P)$ into a valid input for Ising based AQC processors.  
First of all, if we are not worried about the number 
of the extra variables, then   this reduction can be done in  a fairly easy, quick way.  The  idea is
to replace each pair $y_iy_j$ with a new additional variable~$x_{k}$ and add the following expression 
\begin{equation}\label{extraTerm}
M \times \bigg(a_{{1}} \left( x_{{k}}y_{{i}}-x_{{k}} \right) +a_{{2}} \left( x_{{k}}y
_{{j}}-x_{{k}} \right) +a_{{3}} \left( y_{{i}}y_{{j}}-x_{{k}} \right) \bigg) 
 \end{equation}
 as a penalty term (with a large positive coefficient $M$ \footnote{This coefficient is not to be confused with
the ferromagnetic coupling (also denoted $M$ in Section 5), which is a negative large coefficient that maintains the chains. })  to the new function, in order to enforce the equality $x_{k}=y_iy_j$. 
The real numbers  $a_1, a_2,$ and $a_3$ are subject only to $a_{{1}}<-a_{{3}},\, a_{{2}}<-a_{{3}},$ and $a_{{3}}>0$  (accommodating the dynamical ranges of the hardware parameters - See \cite{raouffactorization}).  

~~\\
The connection to toric ideals appears when  we
try to minimize the number of the additional variables
(this is certainly desirable, because additional variables are wasted qubits). This optimization  is  now NP-hard --  See 
 \cite{Boros:2002:PO:772382.772388}. 
Let us  consider the ideal  $\mathcal K_A$   given by   
\begin{eqnarray}\label{idealPairs}
	\mathcal K_A &=& \big \langle  x_{1}        - y_1, \,x_{2} - y_2, \, x_{3} - y_3, \, \cdots, \, x_{m}   -  y_m, \\\nonumber 
	&&\quad x_{k}  - y_{i_1} y_{i_2} ,  \, \mbox { for each pair } (y_{i_1},  y_{i_2}) \mbox { contained in } f \big \rangle,
\end{eqnarray}
where $k$ runs from 1 to $m+n'$, where $n'$ is the total number of such pairs (with $n'+m\leq n$). The configuration matrix  can be readily extracted from the powers of $y$ (See example below). We are interested in computing the toric ideal $\mathcal J_A = \mathcal K_A\cap \mathbb Q[x]$ that gives the algebraic relations between the variables~$x_i$ (in particular, between the  variables~$x_i$ with $i>m$). In fact, the reduced Groebner basis $\mathcal B$ of  $\mathcal K_A$ with respect to the plex order $y\succ x$ has ``two parts": the toric ideal
 $\mathcal J_A$ and a {\it rewriting system} that we use to obtain the minimal  quadratic function.  

   \begin{proposition}
   	Let $\mathcal B$ be the reduced Groebner basis 
	of the ideal $\mathcal K_A$ with respect to the plex order~${y\succ x}$. 
	A minimal reduction of the polynomial function $f$ into a quadratic function can be constructed from 
	the generators of $\mathcal B$ if $f$ is at most quartic and by repeatedly applying 
	this procedure if the degree of $f$ is higher than four. 
 \end{proposition}
 Before illustrating this Proposition on a concrete example, let us note the following remarks:
 
~~\\
(a)  A direct corollary of this Proposition is a re-affirmation that the computation of minimal reductions is NP-hard because
  computing the generators of $\mathcal B$,  given by the kernel $\mathrm{Ker}_{\mathbb Z} A$ in Proposition~\ref{ker},  is  NP-hard.  
  
   ~~\\
(b)   The ideal $\mathcal K_A$ itself is a reduced Groebner basis with  respect to the plex order $x\succ y$; thus,   one can compute the normal form ${\sf NF}_{\mathcal K_A}(f)$ (it makes sense to do so).  If  $f$ is quartic, then the polynomial ${\sf NF}_{\mathcal K_A}(f)$ is quadratic. In general,  repeated application of the same procedure will yield a quadratic  function in polynomial time (by performing Euclidean divisions using the $O(n^2)$ generators of  $\mathcal K_A$). This quadratic function is, however, clearly not minimal. The minimal reduction  is obtained by reversing the plex order (as in the proposition above). 
 
 ~~\\
(c) Multiple choices through different reductions  of the objective function  are desirable. This gives even more possibilities for minor-embeddings with potentially different behaviours of the adiabatic Hamiltonian.

 \begin{example}
 		Consider the cubic polynomial  
	\begin{equation}
		f (y_{{1}},y_{{2}},y_{{3}},y_{{4}},y_{{5}}) = y_{{1}}y_{{2}}y_{{3}}+y_{{1}}y_{{3}}y_{{4}}+y_{{1}}y_{{3}}y_{{5}}+
		y_{{2}}y_{{3}}y_{{5}}+y_{{3}}y_{{4}}y_{{5}},
	\end{equation}
	where our objective is to reduce $f$ into a quadratic polynomial that has the same global minima as $f$. Note that such cubic objective functions
	can be found in {\sf 3Sat} problems.  	
	The configuration  matrix is given. by 
\begin{equation}
A = \left[ \begin {array}{cccccccccccccc} 1&0&0&0&0&1&1&1&1&0&0&0&0&0
\\ \noalign{\medskip}
0&1&0&0&0&1&0&0&0&1&1&0&0&0\\ \noalign{\medskip}
0&0&1&0&0&0&1&0&0&1&0&1&1&0\\ \noalign{\medskip}
0&0&0&1&0&0&0&1&0&0&0&1&0&1\\ \noalign{\medskip}
0&0&0&0&1&0&0&0&1&0&1&0&1&1\end {array} \right]
\end{equation}
where for instance, the sixth column   represents the difference of monomials $x_6- y_1y_2 \in \mathcal K_A$ that one gets from the first monomial of $f$. 
The five first columns represent the differences $x_i-y_i \in \mathcal K_A$ for~{$1 \leq i \leq 5$}.  Calculating the normal form
${\sf NF}_{\mathcal K_A}(f)$ gives the  reduction:
	\begin{equation}\label{naiveReduction}
	{\sf NF}_{\mathcal K_A}(f) = y_{{1}}x_{{10}}+y_{{1}}x_{{12}}+y_{{1}}x_{{13}}+y_{{2}}x_{{13}}+y_{{3}}x_{{14}},
\end{equation} 
where the extra variables  are given by $x_{{10}}=y_{{2}}y_{{3}},\, x_{{12}}= y_{{3}}y_{{4}},\, x_{{13}}=y_{{3}}y_{{5}},\, 
x_{{14}}=y_{{4}}y_{{5}}$. Clearly, ${\sf NF}_{\mathcal K_A}(f)$
is not minimal, and the number of the extra variables  can be reduced. So we compute 
a  Groebner basis for the same ideal $\mathcal K_A$, now, 
with respect to the  plex order $y\succ x$. This basis contains the following polynomials: 
\begin{equation}
y_{{3}}x_{{14}}-y_{{4}}x_{{13}}, \, y_{{1}}x_{{13}}-y_{{5}}x_{{7}},\, 
y_{{1}}x_{{12}}-y_{{4}}x_{{7}},\, y_{{1}}x_{{10}}-y_{{2}}x_{{7}}, 
\end{equation}
which we  think of as a rewriting system, that is a set of replacement rules where, for instance, $y_{{3}}x_{{14}}$ is replaced by  $y_{{4}}x_{{13}}$.
We can then re-expresses the polynomial (\ref{naiveReduction}) into the  minimal reduction that  has only two extra variables, $x_7$ and $x_{13}$. 
 \end{example}

\subsection{Algebraic geometry for graph embeddings}
This subsection discusses the second step in the process of compiling  an optimization problem~$(\mathcal P)$
on an Ising AQC processor (the first being the reduction to QUBOs described in subsection \ref{reduction}) . The subsection
 is a relatively long section,  so we give here a short summary. We begin by recalling the definition of a minor embedding 
which is  a mapping $$\phi:Y\rightarrow X$$ from the space of logical qubits to the space of physical qubits. The logical
qubit $y$ is mapped to a chain (or generally, a connected subtree) of $X$.  We then flip this definition
and introduce an equivalent formulation: the embedding $\phi$ gives  rise to a fiber-bundle $$\pi:X\rightarrow, Y$$  where
the fibers of  the surjection $\pi$ are the chains (connected subtrees) of $X$ given by $\phi$:
$$
	\pi^{-1}(y) = \phi(y).
$$
With this new definition, we  express
 the  surjection $\pi$ equationally: each physical vertex of $X$ is a formal linear combination of the vertices of $Y$, and the different constraints
 on $\phi$ translate into algebraic conditions on the coefficients of these linear combinations. In other words, the set of the fiber bundles  $\pi$ (equivalently the
 set of embeddings $\phi$) is an algebraic variety (of finite cardinality). The theory of Groebner bases can be therefore applied to investigate this variety.  In particular, we  systematically answer the following questions:
 \begin{itemize}
 \item Existence  (or non existence) of embeddings $\phi:Y\rightarrow X$. 
 \item Calculating all embeddings $\phi:Y\rightarrow X$ in a {\it compact form} given by a Groebner basis. 
\item  Counting  all embeddings $\phi:Y\rightarrow X$ without solving any equations.
 \end{itemize}
We do so for any fixed size of the chains. In the last part of the section, we  discuss that many of the embeddings are redundant; that is they are of
the form $\pi\circ \sigma$ with $\sigma$ a symmetry of the hardware graph $X$. This undesired redundancy (which affects the efficiency of the computations) can be removed by expressing
our problem of finding  $\pi$ in terms of the invariants of the symmetry $\sigma$ bringing a nice connection with the  theory of invariants.  Many  calculations are done only once (as long as the hardware doesn't change), and
we illustrate the computational benefit with a simple concrete example.

\subsubsection{Embeddings}\label{Embeddings}
In subsection \ref{reduction},  we have explained how the optimization problem $(\mathcal P)$ 
can be reduced into the quadratic  optimization problem \footnote{
Note that we have used the letter $x$ to denote the binary variables
of the reduced function. In the remainder of the paper, the letter $x$
will be used  to denote the vertices of the hardware graph $X$. We
denote the problem variables (vertices of the logical graph $Y$) by the letter $y$.   
}
  \begin{equation}
	argmin_{(y_0, \cdots,y_{m-1})\in \mathbb B ^m} \,   \sum_{ (y_{i_1}, y_{i_2})\in {\bf Edges}(Y)} J_{i_1 i_2}y_{i_1} y_{i_2}
	+ \sum_{j=0}^{m-1} h_j y_j. 
\end{equation}
This reduction is only the first step in the process of compiling the initial $(\mathcal P)$  on  Ising based AQC processors. The next
step is to map or to embed   the associated logical graph $Y$  into the processor graph $X$.  We have mentioned that this is essentially a sequence of blowups of vertices of $Y$ of high degrees.   We now define the notion of embedding more precisely 
 
\begin{definition}\label{olddef}
Let $X$ be a fixed hardware graph. A \emph{minor-embedding} of the graph $Y$ is a map 
\begin{eqnarray}
	\phi: {\bf Vertices}(Y) \rightarrow \mbox{{\bf connectedSubtrees}}(X)
\end{eqnarray}
that satisfies the following condition for each: $( y_1, y_2)\in {\bf Edges}(Y)$, there exists at least one edge in ${\bf Edges}(X)$ connecting the two subtrees $\phi(y_1)$ and $\phi(y_2)$.   
\end{definition} 
 The condition that each vertex model $\phi(y)$ is a connected subtree of $X$ can be relaxed into $\phi(y)$ is a connected subgraph i.e., $\phi(y)\in\mbox{{\bf connectedSubgraphs}}(X)$. Both cases are considered here. 
 
 ~~\\
In the literature there is another, but equivalent   definition of minor embedding  in  terms of deleting and collapsing the edges of $X$. This   follows from the fact
that,  given a minor-embedding $\phi$, the graph 
$Y$ can be recovered from $X$ by collapsing 
each set $\phi(y)$ (into the vertex $y$) and ignoring (deleting) all vertices of $X$ that are not part of any of the subtrees~$\phi(y)$.

~~\\
Obviously, direct graph embeddings (i.e.,   inclusion graph homomorphisms $Y\hookrightarrow X$) are trivial examples of 
minor embeddings  with  $\phi(y)$ reducing to a vertex in $ {\bf Vertices}(X)$ for all $y\in {\bf Vertices}(Y)$.  Therefore, and for the sake of a simple and clean terminology, we shall use the term embedding instead of minor-embedding throughout the remainder of the paper.

~~\\
Suppose $\phi$ is an embedding of the graph $Y$ inside the graph $X$ as in Definition \ref{olddef}. The subgraph of $X$ given by 
\begin{equation}\label{gmt}
\phi(Y) := \cup_{y\in {\bf Vertices}(Y)}\phi(y)
\end{equation} is called  a $Y$ minor (in graph minor theory). 
In the context of quantum computations, it represents what the quantum processor sees, because it doesn't distinguish between normal qubits and chained qubits. 
 For instance, in examples of Figure \ref{blowup},  $\phi(Y)$ is the induced subgraph defined by the colored
vertices. In other words, the graph  $\phi(Y)$ doesn't   keep track of where each logical qubit
is mapped to.  This information is stored in  the {\it hash} map: 
\begin{equation}
	id\times \phi: {\bf Vertices}(Y)\times {\bf Vertices}(Y) \rightarrow {\bf Vertices}(Y)\times \mbox{{\bf Subtrees}}(X),
\end{equation}
which is used  to {\it unembed} the solution returned by the quantum processor. The numerical 
value of the logical qubit $y$
is the sum mod 2 of its replicates values (a strong ferromagnetic coefficient is used to enforce these replicated values to be equal, i.e., acting like a single qubit). Let us finish this review by mentioning  the existence of heuristics for finding embeddings -- see for instance \cite{DBLP:journals/corr/CaiMR14, Boothby2016} and the references therein. 

\subsubsection{Fiber bundles}\label{FiberBundles}
In this subsection we describe a new computational approach for finding embeddings.  The key point here
is that the set of embeddings is an algebraic variety, that is the set of zeros  of a system of polynomial equations. 
This point becomes clear if  we  think of embeddings as mappings from the space of physical qubits to the space of logical qubits,
which is the {\em opposite} direction of the commonly used definition \ref{olddef}.  Indeed, the
  embedding $\phi$ defines a {\it fiber bundle}: 
\begin{equation}\label{fiberBundle}
\pi:{\bf Vertices}(X)\rightarrow {\bf Vertices}(Y) \cup \{0\}
\end{equation} where the pre-image ({\it fiber}) at
each vertex $y\in {\bf Vertices}(Y)$ is the (vertex set of the) connected subtree $\phi(y)$ of $X$  (per Definition~\ref{olddef}).  The pre-image $\pi^{-1}(0)$ is the set of all physical qubits that are not used (they all project to zero). The reason for mapping unused qubits to zero will become clear soon (when we extend the definition $\pi$ to polynomials).  A direct corollary of this representation, is that   the map $\pi$  has the form: 
\begin{eqnarray}\label{rep2}
	\pi(x_i)   &=& \sum _{j} \alpha_{ij} y_j  \\\nonumber
	&\mbox {with }& \sum _{j} \alpha_{ij} = \beta_i, \quad
	  \alpha_{ij_1} \alpha_{ij_2} = 0,\quad
	  \alpha_{ij} (\alpha_{ij}-1)=0,
\end{eqnarray}
where the binary number $\beta_i$ is equal to one if  the physical qubits $x_i$ is used  and zero otherwise. We write
 $domain(\pi)= {\bf Vertices}(X)$ and $support(\pi)={\bf Vertices}(X^\beta)$ with $X^\beta\subset X$ the  subgraph (\ref{gmt}) defined by $\phi$. 
  The fiber  of the map $\pi$ at $y_j\in {\bf Vertices}(Y)$ is   given by 
  \begin{equation}
  	\pi^{-1}(y_j) = \phi(y_j) = \{ x_i\in {\bf Vertices}(X)|\quad \alpha_{ij}=1 \}.
  \end{equation}
 The conditions on  
  the parameters $\alpha_{ij}$ guarantee that   fibers don't intersect (i.e., $\pi$ is well defined map).

\begin{example}
	Let $X$ and $Y$ be the two graphs depicted in Figure \ref{squareAndTriangle}. An example of the map $\pi$ is defined by $\pi(x_1)=\pi(x_4)=y_1$
	and $\pi(x_2)=y_2$ and $\pi(x_3)=y_3$. 	
\begin{figure}[h]%
    \centering
    \subfloat 
    {{\includegraphics[width=3.5cm]{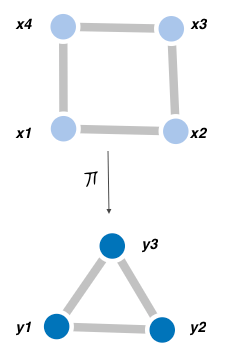} }}%
    \caption{{\footnotesize An example of a fiber bundle of the form (\ref{fiberBundle}).  }}%
    \label{squareAndTriangle}%
\end{figure}

\end{example}

\subsubsection{Finding embeddings}
Before we go deeper in the discussion,  we note that  the number of usable physical qubits  can be constrained: we  
fix  the maximum size of the fibers $\pi^{-1}(y_j)$  to a certain {size~$k \leq \card({\bf Edges}(X))$}. 
This additional size condition can be enforced using: 
\begin{equation}\label{sc}
	\forall j:\quad \sum_{x_i\in {\bf Vertices}(X)} \alpha_{ij} \leq k \quad \mbox { or equivalently } 
	\quad \Pi_{\kappa=1}^k  \left(\sum_{x_i\in {\bf Vertices}(X)}\alpha_{ij} -\kappa \right)=0. 
\end{equation}
Additionally, we have
\begin{equation}
	\forall j:\quad \alpha_{i_1 j}\alpha_{i_2 j}=0,
\end{equation} 
for all pairs $(x_{i_1}, x_{i_2})$ with $d(x_{i_1}, x_{i_2})>k,$ where $d(x_{i_1}, x_{i_2})$ is the size of the shortest chain connecting $x_{i_1}$  and $x_{i_2}$. 

~~\\
The goal of the remainder of this subsection is to translate the two conditions on $\phi$, described in Definition \ref{olddef},   
into a set of polynomial constraints on the parameters $\alpha_{ij}$ and $\beta_i$.  For convenience, we recall the two conditions:
\begin{itemize}
	\item   {\sf Connected Fiber Condition}:  each fiber $\pi^{-1}(y_j)$ of $\pi$ is a  connected subtree of $X.$
	\item   {\sf Pullback Condition}: for each edge $(y_{j_1}, y_{i_2})$ in ${\bf Edges}(Y)$, there exists at least one edge
	in ${\bf Edges}(X)$ connecting the fibers $\pi^{-1}(y_{j_1})$ and $\pi^{-1}(y_{i_2})$.
\end{itemize} 
We start with the first  condition ({\sf Connected Fiber Condition}). We give a conceptually easy characterization. More efficient characterizations can be formulated; particularly, if the tree condition on the fibers is relaxed (i.e.,  $\pi^{-1}(y_j)$ is a connected subgraph of $X$). 
 Let us   introduce the following notations: 
\begin{itemize} 
	\item  $c_k(x_{i_1}, x_{i_2})$ is a chain of size $\leq k$ connecting $x_{i_1}$  and $x_{i_2}$.  Our convention here is
	to define a  chain  as an ordered list of vertices that includes the end points $x_{i_1}$  and $x_{i_2}$, thus, ${\card(C_k(x_{i_1}, x_{i_2}))\leq k+1}$. 
	\item   $\mathcal C_k(x_{i_1}, x_{i_2})$ is the set of all chains of size $\leq k$   connecting $x_{i_1}$  and $x_{i_2}$. \end{itemize} 
Now,  if  two vertices $x_{i_1}$ and $x_{i_2}$ project to $y_j$ (i.e., 
 $\alpha_{i_1j}\alpha_{i_2j}=1$) then there exist a chain $c_k(x_{i_1}, x_{i_2})$ that projects  to $y_j$. This statement is expressed as follows:
\begin{equation} \label{cc0}
	\alpha_{i_1j}\alpha_{i_2j} \times \Pi_{c_k(x_{i_1}, x_{i_2})\in \mathcal C_k(x_{i_1}, x_{i_2})} 
	\left(\Pi_{x_{\ell} \in  c_k(x_{i_1}, x_{i_2})\backslash \{x_{i_1}, x_{i_2} \}} \alpha_{\ell j} -1 \right)=0.
\end{equation}
This condition guarantees only the existence; it doesn't exclude the case when two or more different chains in $ \mathcal C_k(x_{i_1}, x_{i_2})$ 
project to the same $y_j$.  In case  when this is not desirable (that is, when fibers are required to be subtrees of $X$), we need to modify it so that 
one and only one chain projects to $y_j$ (whenever  $x_{i_1}$ and $x_{i_2}$ project to~$y_j$). Thus, instead of \pref{cc0}, 
we impose: 
\begin{equation}\label{cc}
	\alpha_{i_1j}\alpha_{i_2j} \times 
	\left(\sum_{c_k(x_{i_1}, x_{i_2})\in \mathcal C_k(x_{i_1}, x_{i_2})} 
	\Pi_{x_{\ell} \in  c_k(x_{i_1}, x_{i_2})\backslash \{x_{i_1}, x_{i_2} \}} \alpha_{\ell j} -1 \right)=0.
\end{equation}
For each pair of vertices in $\pi^{-1}(y_j)$,
  condition \pref{cc} implies the existence of a unique chain connecting the pair and that is completely contained in the fiber $\pi^{-1}(y_j)$.  Note that,
  the existence of chains  implies that $\pi^{-1}(y_j)$ is connected. 

\begin{proposition}
	Suppose the fiber-bundle $\pi$ given by \pref{rep2} is constrained by 
	the conditions \pref{sc} and \pref{cc}.  Then  
	 fiber $\pi^{-1}(y_j) = \{ x_i\in {\bf Vertices}(X):\, \alpha_{ij}=1 \}$ is a subtree of $X$
	 with size $\leq k$.
\end{proposition}
To prove this statement it suffices to consider three vertices $x_{i_1},x_{i_2},$ and $x_{i_3}$ in $\pi^{-1}(y_j)$
and prove that they cannot form a cycle contained in $\pi^{-1}(y_j)$. Indeed,
if this is  the case, then  $x_{i_1}$ and $x_{i_2}$ are connected
with two different chains of size $\leq k$, which is not possible per conditions \pref{cc}.

~~\\
In case we wish  the fiber
$\pi^{-1}(y_j)$ to be a chain, a preferred minimal structure for the logical qubits, we constrain the degree of each
vertex 
$x_{i_1}$ to be in $\{1, 2\}$, which translates into 
\begin{equation}
-1+\sum_{i_2:\,  (x_{i_1}, x_{i_2}) \in {\bf Edges}(X)}\alpha_{i_1j}\alpha_{i_2j}
\end{equation}
is binary for all  $x_{i_1}\in \pi^{-1}(y_j)$.

~~\\
Let us turn to the {\sf Pullback Condition}, which  states that for each edge $(y_{j_1}, y_{i_2})$ in $Y$ there exists at least one edge connecting
the chains $\phi(y_{j_1})$ and $\phi(y_{i_2})$.   To express this in terms of the parameters $\alpha_{ij}$ and $\beta_i$, we need a few more  constructions:  The map $\pi$ given by the equations  (\ref{rep2}) extends  to  a {\it linear and
multiplicative} map   
\begin{equation}
	\pi:\mathbb Q[{\bf Vertices}(X)] \rightarrow \mathbb Q[{\bf Vertices}(Y)]
\end{equation} by 
\begin{equation}
	\pi(x_{i_1} x_{i_2})  = \pi(x_{i_1}) \pi (x_{i_2})\, \mbox{ and } \pi(a_{i_1}  x_{i_1} + a_{i_2}  x_{i_2}) = 
	a_{i_1} \pi(x_{i_1}) + a_{i_2}  \pi(x_{i_2}),
\end{equation}
for all $a_i\in \mathbb Q.$ The {\it pullback} of  the polynomial $P(x)$ by $\pi$ is  the    polynomial
\begin{equation}
	\pi^*(P)(y)  = P(\pi(x))  \quad \in Q[{\bf Vertices}(Y)].
\end{equation}
In particular, the pullback of the quadratic form $Q_X(x) =\sum_{(x_{i_1}, x_{i_2})\in {\bf Edges}(X)} x_{i_1} x_{j_2}$ by $\pi$ is 
the quadratic form $\pi^*(Q_X)(y)\in \mathbb Q[y]$ given by:
\begin{eqnarray}\nonumber
\pi^*(Q_X)(y)  &=&  \sum_{(x_{i_1}, x_{i_2})\in {\bf Edges}(X)} \pi(x_{i_1}) \pi(x_{i_2})  \\\nonumber
&=& \sum_{(x_{i_1}, x_{i_2})\in {\bf Edges}(X)} \left( \sum_{0\leq j_1< j_2\leq m-1}\left( \alpha_{  i _1 j _1}\alpha_{  i _2 j _2}+\alpha_{  i _1j _2}\alpha_{  i _2 j _1} \right) y_{  j _1}y_{  j _2} + \sum _{j=0}^{ m-1}\alpha_{i_1,j}\alpha_{i_2,j} {y_j}^{2} \right)\\\nonumber
&=& \sum_{0\leq j_1< j_2\leq m-1}   \left(\sum_{(x_{i_1}, x_{i_2})\in {\bf Edges}(X)} \left( \alpha_{  i _1 j _1}\alpha_{  i _2 j _2}+\alpha_{  i _1j _2}\alpha_{  i _2 j _1} \right) \right) y_{  j _1}y_{  j _2}\\
&&  \qquad  + 
 \sum _{j=0}^{ m-1}  \left(\sum_{(x_{i_1}, x_{i_2})\in {\bf Edges}(X)} \alpha_{i_1j}\alpha_{i_2j} \right){y_j}^{2}. 
\end{eqnarray}
Note that the expression $\alpha_{  i _1 j _1}\alpha_{  i _2 j _2}+\alpha_{  i _1j _2}\alpha_{  i _2 j _1}$ is binary. It is equal to one
if  and only if the edge $(x_{i_1}, x_{i_2})$ connects the two fibers $\pi^{-1} (y_{  j _1})$ and $\pi^{-1} (y_{  j _2})$. The sum 
$$
	\sum_{(x_{i_1}, x_{i_2})\in {\bf Edges}(X)} \left( \alpha_{  i _1 j _1}\alpha_{  i _2 j _2}+\alpha_{  i _1j _2}\alpha_{  i _2 j _1} \right) 
$$
gives the number of edges in ${\bf Edges}(X)$ that connect $\pi^{-1} (y_{  j _1})$ and $\pi^{-1} (y_{  j _2})$. 
The  {\sf Pullback Condition} is equivalent to   the fact that this number is strictly non zero
if the pair $\{y_{j_1}, y_{j_2}\}$ is an edges of $Y$. Indeed, this number (a sum of non-negative monomials) is non zero if and only if
there exists a non zero monomial  $\alpha_{  i _1 j _1}\alpha_{  i _2 j _2}$, or equivalently, 
the existence of an edge $(x_{i_1}, x_{i_2})$ connecting the fibers $\pi^{-1}(y_{j_1})$ and $\pi^{-1}(y_{j_2})$. 
\begin{proposition}
	The  {\sf Pullback Condition} is equivalent to the following statement: for each
	$\{y_{j_1}, y_{j_2} \}$ in ${\bf Edges}(Y)$ we have
	\begin{equation}
		\sum_{(x_{i_1}, x_{i_2})\in {\bf Edges}(X)} \left( \alpha_{  i _1 j _1}\alpha_{  i _2 j _2}+\alpha_{  i _1j _2}\alpha_{  i _2 j _1} \right) 
		= 1 + \delta_{j_1j_2}^2,
	\end{equation}
	for some integer $\delta_{j_1j_2}\in \mathbb Z$. 
\end{proposition}
Equations (\ref{rep2}), in addition to the conditions in the previous two propositions define  an algebraic
 ideal $\mathcal I \subset \mathbb Q [\alpha, \, \beta, \delta]$. 
 The zero-locus  of $\mathcal I$ gives all   embeddings of $Y$ (of size $\leq k$) inside the hardware graph $X$. In fact, one has:
 \newpage
  \begin{proposition}
 	Let $\mathcal B$ be a reduced Groebner basis for the ideal $\mathcal I$. The following statements are true:
	\begin{itemize}
	\item A  $Y$ minor  exists if and only if $1\notin \mathcal B$. 
	\item If $\mathcal B$ is computed using the elimination order  $\alpha \succ  \beta \succ \delta$ and
	 $1 \notin \mathcal B$, 
	then the intersection $\mathcal B\cap \mathbb Q[\beta,  \delta]$ gives all subgraphs $X^\beta$ of $X$
	that are minors for~$Y$. The remainder of the reduced  Groebner basis gives the corresponding
	  embedding~${\pi_\beta:X^\beta\rightarrow Y}$. 
	\end{itemize}
 \end{proposition}
 The ``complexity" of the variety $\mathcal V(\mathcal{I})$ is   indicative of the complexity of the topology of the graph $Y$. For instance, if $Y$ consists only of an edge, then $\mathcal V(\mathcal{I})$ is the set of all connected subtrees of size $\leq k$, and if $Y$ is a triangle, then $\mathcal V(\mathcal{I})$ will be the set of 
 all cycles in $X$ with dangling trees at the edges. The dependence of the computational advantage of AQC on the complexity of this variety is an interesting open problem. The dependence on
 the individual points of $\mathcal V(\mathcal{I})$ is discussed in \ref{specGapSection}.

\begin{example}
	Consider the two graphs in Figure \ref{simpleExple}.  In this case, equations (\ref{rep2})
	are given by
  	\begin{eqnarray}
	 && \alpha_{{1,1}}\alpha_{{1,2}}, \, \alpha_{{1,1}}\alpha_{{1,3}}, \, 
\alpha_{{1,2}}\alpha_{{1,3}},
\\
&&
\alpha_{{2,1}}\alpha_{{2,2}},\, \alpha_{{2,1}}\alpha_{{2,3}},\, \alpha_{{2,2}}\alpha_{{2,3}},
\\
&&
\alpha_{{3,1}}\alpha_{{3,2}},\, \alpha_{{3,1}}\alpha_{{3,3}},\, \alpha_{{3,2}}\alpha_{{3,3}},
\\
&&
\alpha_
{{4,1}}\alpha_{{4,2}},\, 
\alpha_{{4,1}}\alpha_{{4,3}},\, \alpha_{{4,2}}
\alpha_{{4,3}},
\\
&&
\alpha_{{5,1}}\alpha_{{5,2}},\, \alpha_{{5,1}}\alpha_{{5,3}},\, \alpha_{{5,2}}\alpha_{{5,3}}, 
	\end{eqnarray}
and 
\begin{eqnarray}\nonumber
&&
\alpha_{{1,1}}+\alpha_{{1,2}}+\alpha_{{1,3}}-\beta_{{1}},
\quad 
\alpha_{{2,1}}+\alpha_{{2,2}}+\alpha_{{2,3}}-\beta_{{2}},
\quad 
\alpha_{{3,1}
}+\alpha_{{3,2}}+\alpha_{{3,3}}-\beta_{{3}},
\\\nonumber
&&
\alpha_{{4,1}}+\alpha_{{4,
2}}+\alpha_{{4,3}}-\beta_{{4}}, 
\quad
\alpha_{{5,1}}+\alpha_{{5,2}}+\alpha_{{
5,3}}-\beta_{{5}}. 
\end{eqnarray}
  \begin{figure}[h] %
    \centering
    \subfloat 
    {{\includegraphics[width=4cm]{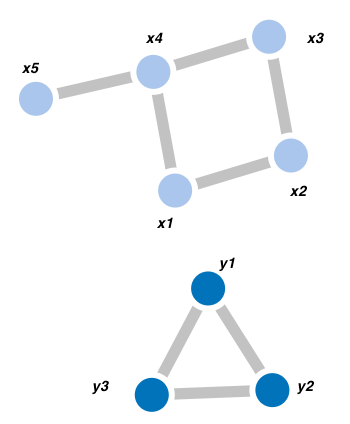} }}%
    \caption[]{{The set of  {\it all} fiber bundles $\pi:X\rightarrow Y$
    defines an algebraic variety. This variety is given by the Groebner basis \pref{GBSimpleExple}.
	}}%
    \label{simpleExple}%
\end{figure}
~~\\
The {\sf Pullback Condition} reads
{\small 
\begin{eqnarray}\nonumber
	&&-1+\alpha_{{4,1}}\alpha_{{5,2}}+\alpha_{{3,1}}\alpha_{{4,2}}+
\alpha_{{1,1}}\alpha_{{2,2}}+\alpha_{{3,2}}\alpha_{{4,1}}+\alpha_{{1,2
}}\alpha_{{2,1}}+\alpha_{{1,2}}\alpha_{{4,1}}+\alpha_{{2,2}}\alpha_{{3
,1}}+\alpha_{{1,1}}\alpha_{{4,2}}+\alpha_{{2,1}}\alpha_{{3,2}}+\alpha_
{{4,2}}\alpha_{{5,1}},
\\\nonumber
&&
-1 +\alpha_{{3,3}}\alpha_{{4,1}}+\alpha_{{1,3}}
\alpha_{{2,1}}+\alpha_{{2,3}}\alpha_{{3,1}}+\alpha_{{4,1}}\alpha_{{5
,3}}+\alpha_{{1,3}}\alpha_{{4,1}}+\alpha_{{1,1}}\alpha_{{2,3}}+\alpha_
{{4,3}}\alpha_{{5,1}}+\alpha_{{2,1}}\alpha_{{3,3}}+\alpha_{{3,1}}
\alpha_{{4,3}}+\alpha_{{1,1}}\alpha_{{4,3}},
\\\nonumber
&& 
-1 +\alpha_{{3,3}}\alpha_{{4,2
}}+\alpha_{{1,2}}\alpha_{{2,3}}+\alpha_{{1,2}}\alpha_{{4,3}}+\alpha_{{
1,3}}\alpha_{{2,2}}+\alpha_{{1,3}}\alpha_{{4,2}}+\alpha_{{2,3}}\alpha_
{{3,2}}+\alpha_{{2,2}}\alpha_{{3,3}}+\alpha_{{4,2}}\alpha_{{5,3}}+
\alpha_{{3,2}}\alpha_{{4,3}}+\alpha_{{4,3}}\alpha_{{5,2}}.  \nonumber
\end{eqnarray}
Finally, the {\sf Connected Fiber Condition} is given by
\begin{eqnarray}\nonumber
&&
-\alpha_{{1,1}}\alpha_{{2,1}}\alpha_{{5,1
}},-\alpha_{{1,1}}\alpha_{{3,1}}\alpha_{{5,1}},-\alpha_{{1,2}}\alpha_{
{2,2}}\alpha_{{5,2}},
-\alpha_{{1,2}}\alpha_{{3,2}}\alpha_{{5,2}},-
\alpha_{{1,3}}\alpha_{{2,3}}\alpha_{{5,3}},-\alpha_{{1,3}}\alpha_{{3,3
}}\alpha_{{5,3}}
\\\nonumber
&&
-\alpha_{{2,1}}\alpha_{{3,1}}\alpha_{{5,1}},-\alpha_{
{2,1}}\alpha_{{4,1}}\alpha_{{5,1}},-\alpha_{{2,2}}\alpha_{{3,2}}\alpha
_{{5,2}},-\alpha_{{2,2}}\alpha_{{4,2}}\alpha_{{5,2}},-\alpha_{{2,3}}
\alpha_{{3,3}}\alpha_{{5,3}},-\alpha_{{2,3}}\alpha_{{4,3}}\alpha_{{5,3
}},
\\\nonumber
&&
\quad
\alpha_{{2,1}}\alpha_{{5,1}},\alpha_{{2,2}}\alpha_{{5,2}},
\alpha_{{2,3}}\alpha_{{5,3}}.
\end{eqnarray}	
}
A part of the reduced  Groebner basis of the resulted system is given by 
\begin{eqnarray}\label{GBSimpleExple}\nonumber
\mathcal B &=&   \left\{ \beta_{{1}}-1,\beta_{{2}}-1,\beta_{{3}}-1,\beta_{{4}}-1, 
 \beta^2_i-\beta_i,\, \alpha_{ij}^2 -\alpha_{ij}, \,\right.
\\[3mm]\nonumber
&&
\alpha_{{1,2}}\alpha_{{1,3}},\alpha_{{1,2}}\alpha_{{3,2}},
\alpha_{{1,3}}\alpha_{{3,3}},\alpha_{{2,2}}\alpha_{{2,3}},\alpha_{{2,2
}}\alpha_{{4,2}},\alpha_{{2,2}}\alpha_{{5,2}},\alpha_{{2,3}}\alpha_{{4
,3}},\alpha_{{2,3}}\alpha_{{5,3}},\alpha_{{3,2}}\alpha_{{3,3}}, \alpha_{{4,2}}\alpha_{{4,3}},
\\[3mm]\nonumber
&& 
\alpha_{{4,2}}\alpha_{{5,3}},
\alpha_{{4,3}}\alpha_{{5,2}},\alpha_{{5,2}}\alpha_{{5,3}},\alpha_{{4,2}}\alpha_{{5,2
}}-\alpha_{{5,2}},\alpha_{{4,2}}\beta_{{5}}-\alpha_{{5,2}},\alpha_{{4,
3}}\alpha_{{5,3}}-\alpha_{{5,3}}, 
\\\nonumber
&&
\quad \quad \vdots\\\nonumber
&&
-\alpha_{{2,2
}}\alpha_{{5,3}}-\alpha_{{3,2}}\alpha_{{5,3}}+\alpha_{{1,2}}\beta_{{5}
}+\alpha_{{2,2}}\beta_{{5}}+\alpha_{{3,2}}\beta_{{5}}+\alpha_{{3,3}}
\beta_{{5}}+\alpha_{{5,2}}+\alpha_{{5,3}}-\beta_{{5}}\left.\right \}. 
\end{eqnarray}
 In particular, the intersection $\mathcal B\cap \mathbb Q[\beta] =(\beta_{{1}}-1,\beta_{{2}}-1,\beta_{{3}}-1,\beta_{{4}}-1, {\beta_{{5}}}^{2}-\beta_{{5}})$ gives
the two $Y$ minors (i.e., subgraphs $X^\beta$) inside $X$.  
 The remainder of $\mathcal B$ gives
the explicit expressions of the corresponding mappings.
\end{example}

\subsubsection{Counting embeddings without solving equations}
The number of zeros of an ideal $\mathcal I\subset \mathbb Q[x_0,\cdots, x_{n-1}]$ can be  determined  without solving
any equation in  $\mathcal I$.  
  This is done using {\it staircase diagrams}, as follows. To each polynomial in $\mathcal I$
we assign a point in the Euclidean space $\mathbb E^{n}$ given by the exponents of its leading term (with respect to the given monomial order). Figure \ref{stairecase} depicts three  staircase diagrams. 

  \begin{figure}[!h]
    \centering 
    {{\includegraphics[width=16cm]{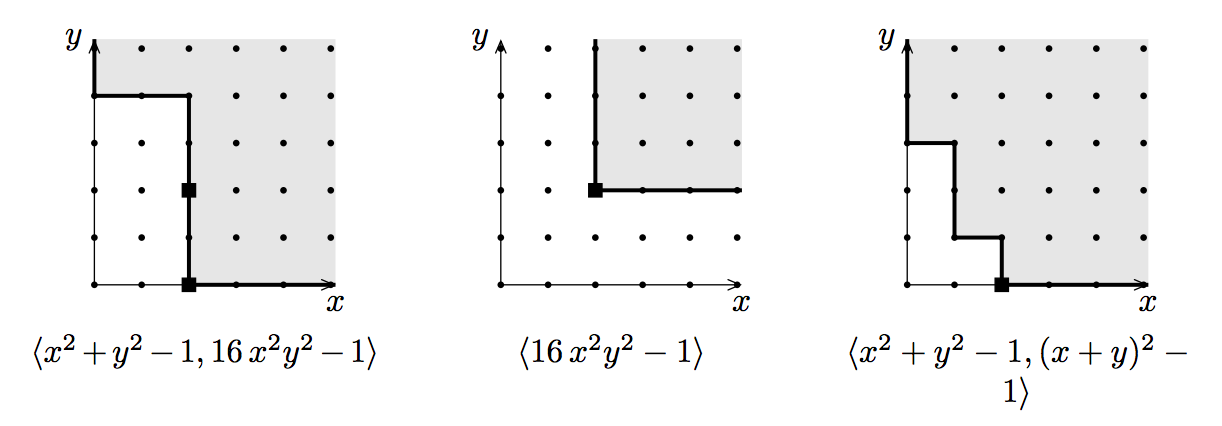} }} 
    \caption{ Staircase diagrams of three ideals in $\mathbb Q[x, y]$. For these examples, the staircase diagrams 
    are the same for the monomial orders lex, deglex, and  degrevlex.}
    \label{stairecase}
  \end{figure}
 
 \begin{proposition}
 The ideal $\mathcal I\subset \mathbb Q[x_0,\cdots, x_{n-1}]$ is zero dimensional if and only if 
 the number of points under the shaded region of its staircase is finite, and this number
 is equal to the dimension of the quotient $Q[x_0,\cdots, x_{n-1}]/\mathcal I$, that is, the number of zeros of $\mathcal I$. 
 \end{proposition}
 One can see that the number of zeros of the three ideals in Figure \ref{stairecase}  (left to right) are $8, \, \infty$ and 4 respectively.  
 
 ~~\\
 The application of this construction to the problem of counting all  embeddings ${\pi: X\rightarrow Y}$ is obvious. The ideal $\mathcal I$ is given by the different requirements on the coefficients $\alpha_{ij}$ of the map $\pi$ as discussed previousely. 
 Note that the dimension of $Q[\alpha, \beta, \delta]/\mathcal I$  cannot be infinite because there is (if any) only finite number of possible embeddings.  An example is depicted in Figure \ref{stairecaseExample}. 
  \begin{figure}[!h]
    \centering 
    {{\includegraphics[width=4cm]{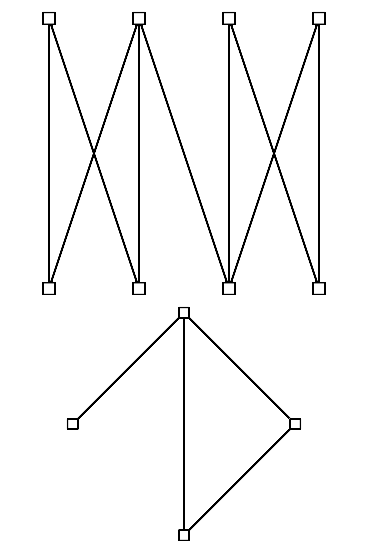} }} 
    \caption{{\small There are 360   embeddings, with chains of size at most 2,  for the bottom graph into the upper graph.}}
    \label{stairecaseExample}
  \end{figure}
  
\subsubsection{Symmetries and invariant coordinates}\label{symmetries}
When determining the  surjections $\pi$  (or equivalently, the   embeddings $\phi)$,  many of the  solutions are  redundant: they are    of the form $\pi\circ \sigma$ with $\sigma \in {\bf Aut}(X)$.  This is   not desirable because it affects the efficiency of the computations. 
 In this subsection, we   discard  this redundancy by expressing  the problem of finding the  fiber bundles $\pi$ in a canonical form, using  the invariants of
the symmetry (automorphism)~$\sigma$.   Pictorially, this  canonical form is obtained by folding the hardware graph $X$ along the symmetry axis.

~~\\
The tools and concepts that we have used here are rooted  in {\it classical invariant theory}; a reference to this fascinating subject is the excellent book \cite{olver_1999}. The first of these concepts is the notion of invariance:
An {\it invariant}  of a symmetry $\sigma \in Aut(\mathcal X)$ of an object $\mathcal X$  is a real-valued 
function $I: \mathcal X\rightarrow \mathbb R$  that  satisfies
\begin{equation}
I(\sigma x)=I(x)
\end{equation} for all $x\in \mathcal X$;  the function
$I$ is invariant of a subgroup $G\subseteq Aut(\mathcal X) $ if and only it is invariant of all $\sigma \in G$. Because
the sum and the product of invariants are again invariants, 
the set of all invariants of  $G$ forms an  algebra (the {\it algebra of $G-$invariants}), denoted $\mathcal A^G$, and is a subalgebra
of the algebra of real-valued functions over $\mathcal X$. The algebra of $G-$invariants $\mathcal A^G$ can be very large. In practice, we don't compute the entire $\mathcal A^G$,  but instead we compute a {\it complete set} of invariants that is a maximal set of invariants that are
 functionally  independent (or algebraically   independent in the context of polynomials). Any other $G-$invariant can be written as a function
 thereof.    
 A useful construction for finite groups is the Reynolds symmetrization operators
\begin{equation}
  \varsigma  = \sum_{\sigma\in G}  \sigma,  
\end{equation}
which projects the algebra of functions over $\mathcal X$   to the algebra $\mathcal A^G$.  In other words, for each
function $f: \mathcal X\rightarrow \mathbb R$, the function $\varsigma f: \mathcal X\rightarrow \mathbb R$
is a $G-$invariant (and  if $f$ is already invariant then $\varsigma f=f$).  This yields a procedure for constructing invariants
of $G$.   Furthermore, a set of invariants is functionally independent if their Jacobian matrix has full rank.

~~\\
In our context of mapping binary optimization problems into  quantum hardwares, the space~$\mathcal X$ is   the hardware graph $X$, in which   case symmetries are permutations of the vertices.  For instance, in the case of  the example  of Figure \ref{folding1}, the 
rotation around the edge $(x_1, x_2)$ is a symmetry. Additionally, the functions $x_1, x_2, x_3+x_4, x_3x_4,  x_5+x_6,$ and 
$x_5x_6$ are invariants. They form
a  complete set of invariants for this symmetry \footnote{In general, if $\sigma$ is an elementary permutation 
that exchanges the two nodes $x_{i_1}$ and $x_{i_2}$ and leaves the rest of the nodes invariant, then the $n$ functions
$x_{i_1} + x_{i_2}, x_{i_1}x_{i_2}$ and $x_i$ for $i\in \{1,\cdots, n\}\backslash \{i_1, i_2\}$ form a   complete set of invariants.}.  
Figure \ref{folding1} (right) gives the canonical representation of the graph $X$ with respect to  its symmetry. 
 

  \begin{figure}[h]  %
    \centering
    \subfloat 
    {{\includegraphics[width=8cm]{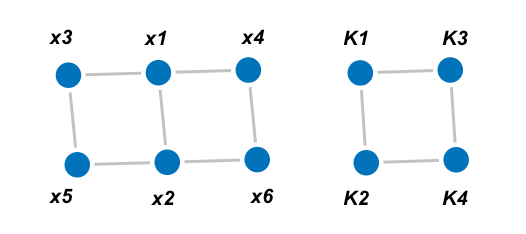} }}%
    \caption[]{{ The hardware graph $X$ (right) is symmetric along the edge $x_1x_2$.  The nodes of the folding (right)
      correspond to the 4 functionally independent  invariants of this symmetry: $K_1=x_1,$ $K_2=x_2,$ $ K_3=x_3+x_4 $ and
      $K_4 = x_5 +x_6$.  
	}}%
    \label{folding1}%
\end{figure}

~~\\ 
In the remainder of this section we  explain how this canonical representation can be determined in general.  To do so,  the notion of
 ``algebra of functions over a graph $X$" needs to be made precise.  Indeed, this algebra is  the quotient ring: 
\begin{equation}
	\mathcal A_X := \mathbb Q[{\bf Vertices} (X)]/ \langle \mbox{monomials } x_{i_1} x_{i_2} | \quad (x_{i_1}, x_{i_2})\notin  {\bf Edges} (X)  \rangle. 
\end{equation} 
Readers familiar with the notion of  coordinates rings might notice that $\mathcal A_X$ is the ``coordinates ring" of the complement graph of $X$. 
Now, suppose   $G$ is a subgroup of $ {\bf Aut}(X)$ (which is itself a subgroup of the symmetric group $S^n$). The set of $G$-invariants  forms  a subalgebra~$\mathcal A^G_X $ of the algebra of functions $\mathcal A_X$.  Folding the graph $X$ along $G$ consists then of applying the following
procedure:
{\sf 
\begin{itemize}
	\item[] {\bf input}: A graph $X$ and a complete set of $G-$invariants $\{I_0(x), \cdots, I_{r-1}(x)\}$
	\item[] {\bf output:} The folding of $X$ along $G$ 
	\item[1] define the system $$\mathcal S=   \{f-Q_X(x)\}\cup \{K_0-I_0(x), \cdots, K_{r-1}-I_{r-1}(x)\} \cup  \{x_{i_1} x_{i_2} | \quad \{ x_{i_1}, x_{i_2} \}\notin  {\bf Edges} (X) \}.$$
	The system $\mathcal S$ is a subset of the extended polynomial ring $\mathbb Q[x_0, \cdots,x_{n-1},K_0, \cdots, K_{r-1}, f]$.
	\item[2] compute a Groebner basis $\mathcal B$ for $\mathcal S$ with respect to the monomial order  
	$$	
		[x_0, \cdots,x_{n-1}] \succ f \succ [K_0, \cdots, K_{r-1}].
	$$
	The intersection $\mathcal B\cap \mathbb Q[K_0, \cdots, K_{r-1}, f]$ gives folding of $X$ along $G$.  
\end{itemize}
}

~~\\
The use of Groebner bases in this procedure is not a necessity -- it is more for conciseness and beauty of the formulation. 
Another way to do the same task without computing any Groebner basis is by solving the system $\{K_i-I_i\} $
with respect to the variables $x_1,\cdots, x_n$, replacing the solution in $X$ and then evaluating   on 
the ideal
\begin{equation}{\langle \mbox{monomials } x_{i_1} x_{i_2} | \quad ( x_{i_1}, x_{i_2})\notin  {\bf Edges} (X)  \rangle}.
\end{equation}
Let us illustrate this procedure on the example depicted in Figure \ref{folding1} above. The system $\mathcal S$ is given by the polynomials 
\begin{eqnarray}
	&& {K_1}-x_{{1}},\, {K_2}-x_{{2}},\, {K_3}-x_{{3}}-x_{{4}},\, {K_4}-x_{{5}}-x_{{6}},\, {K_5}-x_{{3}}x_{{4}},\, {K_6}-x_{{5}}x_{{6}},\, \\
	&& f-x_{{3}}x_{{1}}-x_{{1}}x_{{4}}-x_{{3}}x_{{6}}-x_{{4}}x_{{5}}-x_{{1}}x_{{2}}-x_{{2}}x_{{6}}-x_{{5}}x_{{2}},\, \\
	&& x_{{3}}x_{{4}},\, x_{{5}}x_{{6}},\, x_{{4}}x_{{6}},\, x_{{3}}x_{{5}},\, x_{{1}}x_{{5}},\, x_{{1}}x_{{6}},\, x_{{2}}x_{{4}},\, x_{{2}}x_{{3}}. 
\end{eqnarray}
We compute a Groebner basis $\mathcal B$ as in the procedure above.
In this case, the intersection $\mathcal B\cap \mathbb Q[K_1,\cdots, K_6, f]$ is the set of polynomials: 
\begin{equation}
{K_6},\quad {K_5},\quad {K_4}\,{K_1},\quad {K_3}\,{K_2},\quad f-{K_1}\,{K_2}-{K_1}\,{K_3}-{K_2}\,{K_4}-{K_3}\,{K_4}.
\end{equation}
The last polynomial gives the rectangular graph of Figure \ref{folding1} with nodes $K_1, K_2, K_3,$ and~$K_4$. The example of
Figure \ref{folding2} is treated similarly.

%
%
%
%

  \begin{figure}[h]  %
    \centering
    \subfloat 
    {{\includegraphics[width=7cm]{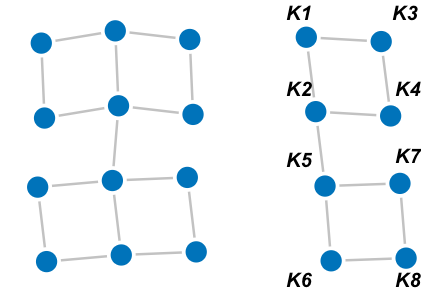} }}%
    \caption[]{{ Folding of the hardware graph $X$ (left) along its symmetry axis. The nodes of the folding (right) are functionally independent  invariants of the symmetry. 
	}}%
    \label{folding2}%
\end{figure}


~~\\
We finish this subsection with an example illustrating  how the embedding is found once redundancy is removed. 
  \begin{example}      
  	Consider the two graphs $X$ and $Y$ of  Figure \ref{simpleExple}. The  quadratic form of $X$ is:
	\begin{equation}
		Q_X(x) = x_{1}x_{2}+x_{2}x_{3}+x_{3}x_{4}+x_{1}x_{4}+x_{4}x_{5}. 
	\end{equation}
	Exchanging the two nodes $x_1$ and $x_3$ is a symmetry for $X,$ and the quantities $K=x_1 + x_3,  \, x_2, \,  x_4,$ and  $x_5$ are invariants of this symmetry.	In terms of these invariants, the quadratic function $Q_X(x)$,  takes the simplified form: 
	\begin{equation}
		Q_X(x, K) = K x_{2}+K x_{4} +x_{4}x_{5},  
	\end{equation}
	which shows (as expected) that graph $X$ is folded into a chain (given by $[x_2, K, x_4, x_5]$). 
	The surjective homomorphism $\pi:X\rightarrow Y$  now takes  the form
	\begin{system}
		K &=& \alpha_{01}y_1 +\alpha_{02}y_2 +\alpha_{03}y_3.\\
		x_i &=&  \alpha_{i1}y_1 +\alpha_{i2}y_2 +\alpha_{i3}y_3 \mbox { for } i=2, 4, 5. 
	\end{system}%
	The coefficients $\alpha_{ij}$ for $i=2, 4, 5$ are constrained as usual.  The coefficients $\alpha_{0j}$ are 
	binary subject to $\sum_{j}\alpha_{0j}\leq 2$. In general, if $K=\sum_{i\in I} x_i$ where
	$d(x_{i_1}, x_{i_2})> k$ for all ${i_1}, {i_2}\in I$, then we have $\sum_{j}\alpha_{0j}\leq \mathrm{card}(I)$. The number $k$ is the maximum allowed size of the chains.  The table below compares the computations of the surjections $\pi$  with and without the use of invariants:

	\begin{center}
	\begin{tabular}{l|l|l}
						  	 					& original coordinates 	&  invariant coordinates\\\hline
		Time for computing a  Groebner basis (in secs)	& 0.122 			& 0.039\\
		Number of defining equations					& 58				& 30\\
		Maximum degree in the defining equations  		&  3				& 2\\
		Number of variables	in the defining equations 		& 20				&12\\
		Number of solutions							& 48				& 24\\
		\hline
 	\end{tabular}  
	\end{center}
  	
~~\\
In particular, the number of solutions is down
 to 24, that is, four (non isomorphic) minors times
 the six symmetries of the logical graph $Y$. 
  \end{example} 
Let us conclude this section by noting  that symmetries of the problem graph  can also be considered; however, this is problem dependent and needs to be redone  if the problem is changed (unlike the hardware graph $X$, which is fixed). 

\subsection{Analytical dependence of the spectral gap of the adiabatic Hamiltonian on the points of the embedding variety}\label{specGapSection}
Consider a hardware graph $X$ and a problem graph $Y$. 
This section
discusses 
 the dependence of the computational complexity of AQC on the choice of the embedding of $Y$ inside $X$.
 
 ~~\\One of the key findings of this paper is that  the set of  embeddings $\pi:X\rightarrow Y$ is an algebraic variety i.e., a geometrical object with a coordinate system given by
a reduced  Groebner~$\mathcal B$. Our problem translates into   determining
 the algebraic dependence of the computational complexity of AQC on the coordinates of the variety $\mathcal V(\mathcal B)$.   That is, determining the algebraic dependence of the spectrum 
of the adiabatic Hamiltonian 
\begin{equation}
H(t)=\alpha(t) H_{initial} +\beta(t) H_{(\mathcal P)}
\end{equation}
 on  the coordinates of the variety $\mathcal V(\mathcal B)$. One way to proceed is to
  obtain
 the most general expression of the quadratic form of the~$Y$  minor, which we denote by $\tilde \phi(Q_Y)(x)$, in terms of the parameters $\alpha_{ij},$ and $\beta_i$ determined by $\mathcal B$. 
 This is given by the proposition below, which uses the following observation. Each variable $x_i$ can be represented by a row vector $(\alpha_{i1}, \cdots, \alpha_{im})\in \{0,1\}^m$. In this case,
 two physical qubits $x_{i_1}$ and $x_{i_2}$ are chained if and only if their dot product $\sum_{j}  \alpha_{i_1j}  \alpha_{i_2j} $
is not equal to zero. Similarly, the qubit $x_{i}$ is selected  (i.e., $x_i\in support(\pi)$) if and only if $\sum_{j}  \alpha_{ij}$  is not zero. 

\begin{proposition}\label{specGap}
Given a hardware graph $X$ and a problem graph $Y$. Let $\mathcal B$
 denote the reduced  Groebner basis that gives the set of  embeddings $\pi:X\rightarrow Y$. The general
 form of the quadratic form of the $Y$ minor is given by 
 \begin{eqnarray}
 \tilde \phi(Q_Y)(x) &= &\sum_{x_{i_1}x_{i_2}\in {\bf Edges}(X)} 
 {\sf NF}_{\mathcal B} \left\{ \left(\sum_{j}  \alpha_{i_1j} \right) \left(\sum_{j}  \alpha_{i_2j}\right) \right\} {x_{i_1}x_{i_2}} \\[3mm]\nonumber
  && \quad + \quad    
	  M \times {\sf NF}_{\mathcal B} \left\{ \sum_{j}  \alpha_{i_1j}  \alpha_{i_2j} \right\}  {(1-2x_{i_1}) (1-2x_{i_2})},
 \end{eqnarray}
with $M$ being one (or more)  strong ferromagnetic coupling that  maintains the chain.  Therefore,   the problem Hamiltonian 
is 
\begin{equation}
	H_{(\mathcal P)}(\mathcal B) =   \sum_ {\mathrm i=(\mathrm i_1, \cdots, \mathrm i_n)\in \{0, 1\}^n  }{\sf NF}_{\mathcal B} \left\{ \tilde \phi(Q_Y)(\mathrm i)  \right\} |\mathrm i\rangle \langle \mathrm i|
\end{equation}
given in the computational basis of the Hilbert space ${\mathbb C^2}^{\otimes n}.$
\end{proposition}
Recall that the notation ${\sf NF}_{\mathcal B}(f)$ stands for the normal formal of the polynomial $f$ with respect to  ${\mathcal B}$. This is reviewed in the mathematical background section. We conclude with a simple example. 
\begin{example}
	Consider the two graphs given by the quadratic functions $Q_X(x)=x_{{1}}x_{{2}}+x_{{2}}x_{{3}}$ and $Q_Y(y)=y_{{1}}y_{{2}}$. In this case, the reduced  Groebner basis (computed using tdeg order) is given by 
	\begin{eqnarray}\nonumber
	&&\beta_{{2}}-1, 
	\quad (\beta_{{1}} -1) (\beta_{{3}} -1), 
	\quad {\beta_{{3}}}^{2}-\beta_{{3}}, 	
	\quad {\beta_{{1}}}^{2}-\beta_{{1}}, \, \\[3mm]\nonumber
	&&
	\alpha_{{1,2}}\alpha_{{3,2}},\quad
	\alpha_{{2,1}}+\alpha_{{2,2}}-1,\quad
	\alpha_{{3,1}}+\alpha_{{3,2}}-\beta_{{3}},\quad
	\alpha_{{1,1}}+\alpha_{{1,2}}-\beta_{{1}},\quad
	\alpha_{{1,2}}\beta_{{1}}-\alpha_{{1,2}},\quad
	\alpha_{{3,2}}\beta_{{3}}-\alpha_{{3,2}},
	\\[3mm]\nonumber
	&&\alpha_{{1,2}}\beta_{{3}}+1+\alpha_{{2,2}}\beta_{{3}}-\alpha_{{1,2}}-\alpha_{{2,2}}-\beta_{{3}},\quad
	\alpha_{{3,2}}\beta_{{1}}-\alpha_{{2,2}}\beta_{{3}}+\alpha_{{1,2}}+\alpha_{{2,2}}-\beta_{{1}},\\[3mm]\nonumber
&&\alpha_{{2,2}}\beta_{{1}}+1+\alpha_{{2,2}}\beta_{{3}}-\alpha_{{1,2}}-2\,\alpha_{{2,2}}-\alpha_{{3,2}}, \quad
\alpha_{{1,2}}\alpha_{{2,2}}+1+\alpha_{{2,2}}\alpha_{{3,2}}-\alpha_{{1,2}}-\alpha_{{2,2}}-\alpha_{{3,2}},\\[3mm]
&& {\alpha_{{1,2}}}^{2}-\alpha_{{1,2}}, \quad \quad {\alpha_{{3,2}}}^{2}-\alpha_{{3,2}}, \quad {\alpha_{{2,2}}}^{2}-\alpha_{{2,2}}. 
	\end{eqnarray}
	The first four polynomials give the reduced Groebner basis $\mathcal B\cap \mathbb Q[\beta_1, \beta_2,\beta_3]$, which gives the different domains for the projection $\pi$.  
	The general form of $Y$ minor is given  by
	\begin{equation}
		\tilde \phi(Q_Y)(x) =\beta_{{1}}x_{{1}}x_{{2}}+\beta_{{3}}x_{{2}}x_{{3}}+M \left( -1+\beta_
{{1}}+ \gamma\right)(-2x_{{1}}+1) (-2x_{{2}}+1) + M \left( \beta_{{3}}-\gamma \right) (1-2x_{{ 2}})(1-2x_{{3}}),
	\end{equation}
		with $\gamma = \alpha_{{3,2}} +\alpha_{{2,2}}\beta_{{3}}-2\alpha_{{2,2}}\alpha_{{3,2}}$. 
	
\end{example}
We envision two additional applications of the previous proposition. First, as a sandbox, it can help in building intuition and providing fruitful directions for further investigation through studying small instances
in a systematic, principled and comprehensive manner. The second application is specific to the case when one has a structured class of problems where the scaling follows a certain formulaic description. The general pattern of the adiabatic behaviour might emerge from studying small problem instances.
\section{Designing Ising Quantum Architectures}
A critical milestone in the development of AQC is the design of 
Ising architectures that can be physically realized and are capable of solving some important, hard problems -- for instance, designing architectures optimized for a particular  class of problems in machine learning.  
This is formalized as  
designing hardware graphs $X$ that satisfy the following: 
\begin{itemize}
\item The degree of $X$ cannot exceed a limited degree $d$ (imposed by current manufacturing limitations). 
\item $X$ contains a minor for each graph $Y \in \mathcal{Y}$,
where $\mathcal{Y}$   represents
a class of problems of interest.  
\item Each $Y$ minor is explicitly computable. 
\end{itemize}
This problem as described was  posed  in  \cite{Choi2011}, where the following nomenclature was introduced:
\begin{definition}
	Let $\mathcal Y$ be a  family of graphs. A graph $X$ is called $\mathcal Y-$minor universal if for any graph
	$Y\in \mathcal Y,$ there exists a minor embedding of $Y$ in $X$. 
\end{definition}
At this point, the reader might anticipate that our approach
is able to produce  such $\mathcal Y-$minor universals. Indeed, it is easy to see
that the first requirement translates into the condition $\sum_j q_{ij}\leq d$, where
 $(q_{ij})_{1\leq i, j\leq n}$ is the unknown adjacency matrix of $X$.  Additionally, if 
 the family $\mathcal Y$ is given by a finite number of graphs $Y_\mu$ (where $\mu$ belongs to  a finite range), then for each graph $Y_\mu$, we define the transformation
\begin{equation}
	\pi^\mu (x_i)= \sum_{y_i\in V(Y_i)} \alpha^\mu_{ij} y_j,
\end{equation} 
where the binary coefficients are subject to the conditions (\ref{rep2}) for each index $\mu$.   These conditions, in addition to the pullback and  connected fiber conditions  for all $\mu$ as well as the degree condition above,  form a system of polynomials $\mathcal L\subset \mathbb Q[ \alpha^\mu, q]$ that has all information needed to determine 
the coefficients~$q_{ij}$.  More precisely, we have   
\begin{proposition}\label{YminorUniversal} Let $\mathcal B$ be a reduced  Groebner basis for the system $\mathcal L$ with respect to the elimination order 
	$ \{\alpha_{ij}^\mu\} \succ  \{q_{ij}\}$. The following statements are true:
	\begin{itemize}
		\item the family of graphs $\mathcal Y=\{Y_\mu\}$  admits a $\mathcal Y-$minor universal graph of size $n$
		 if and only if $1\notin \mathcal B$ (the choice of the ordering used is not relevant for this statement).
		\item if $1\notin \mathcal B$, the set of all $\mathcal Y-$minor universal graphs of size $n$ is given by the intersection $\mathcal B\cap \mathbb Q[q_{ij}]$.
		\item  if $1\notin \mathcal B$, the   embeddings $\pi^\mu$ (i.e., the coefficients $ \alpha^\mu_{ij}$) 
		are also given by $\mathcal B$ (as functions of the~$q_{ij}$). 
	\end{itemize} 	
\end{proposition}
%
%
%
It is easy to see that 
Proposition \ref{YminorUniversal} also holds when the finite family $\mathcal Y=\{Y_\mu\}$ is replaced with a parametrized family of graphs provided the conditions on the parameters are algebraic: these additional conditions
are added to the ideal  $\mathcal L$, which is now a subset of the polynomial ring $\mathbb Q[ \alpha^\mu, q, \mu]$. In this case, we compute a reduced  Groebner basis for the system $\mathcal L$ with respect to the elimination order $ \{\alpha_{ij}^\mu\} \succ  \{q_{ij}\}  \succ  \{\mu \}$. 

~~\\
Another possible extension is by combining Proposition \ref{YminorUniversal}   with Proposition \ref{specGap}. 
The latter gives the anlaytical dependance of the spectral gap of the adiabatic Hamiltonian on the the variety $\mathcal V(\mathcal B) =\{\pi:X\rightarrow Y \}$.  In  Proposition \ref{specGap}, we consider a family $\mathcal Y$  of graphs (instead of one graph $Y$). In that case, we  obtain the dependence of the spectral gap of $H(t)$ on the choice of the $\mathcal Y$ minor universal as well as the different corresponding embeddings. 

\section{Concluding Remarks}
In this paper, we developed a novel algebraic geometry framework to solve integer polynomial optimization using adiabatic quantum computing. Our approach represents the first fully systematic translator
for AQC in which the intricate steps in the compiler process are codified algorithmically.  This approach can also serve as a test-bed to design scalable compilers (by empirically testing the performance of various embeddings that differ on chain length, number of physical qubits used and other features) and Ising architecture design.  

\section{Acknoledgements}
We thank Jesse Berwald, Denny Dahl and Steve Reinhardt from D-Wave systems 
who provided insight and expertise.   We  also thank 
Eleanor G. Rieffel, Bryan O'Gorman, Davide Venturelli  and NASA QuAIL team members   for comments and feedback.

\newpage

 \bibliography{c}

\newcommand{\etalchar}[1]{$^{#1}$}
\def\cprime{$'$} \def\cprime{$'$}
\providecommand{\bysame}{\leavevmode\hbox to3em{\hrulefill}\thinspace}
\providecommand{\MR}{\relax\ifhmode\unskip\space\fi MR }
\providecommand{\MRhref}[2]{%
  \href{http://www.ams.org/mathscinet-getitem?mr=#1}{#2}
}
\providecommand{\href}[2]{#2}
\begin{thebibliography}{AvDK{\etalchar{+}}04}

\bibitem[AE99]{Avron1999}
Joseph~E. Avron and Alexander Elgart, \emph{Adiabatic theorem without a gap
  condition}, Communications in Mathematical Physics \textbf{203} (1999),
  no.~2, 445--463.

\bibitem[AvDK{\etalchar{+}}04]{1366223}
D.~Aharonov, W.~van Dam, J.~Kempe, Z.~Landau, S.~Lloyd, and O.~Regev,
  \emph{Adiabatic quantum computation is equivalent to standard quantum
  computation}, 45th Annual IEEE Symposium on Foundations of Computer Science,
  Oct 2004, pp.~42--51.

\bibitem[Bar82]{0305-4470-15-10-028}
F~Barahona, \emph{On the computational complexity of ising spin glass models},
  Journal of Physics A: Mathematical and General \textbf{15} (1982), no.~10,
  3241.

\bibitem[BF28]{Born1928}
M.~Born and V.~Fock, \emph{Beweis des adiabatensatzes}, Zeitschrift f{\"u}r
  Physik \textbf{51} (1928), no.~3, 165--180.

\bibitem[BH02]{Boros:2002:PO:772382.772388}
Endre Boros and Peter~L. Hammer, \emph{Pseudo-boolean optimization}, Discrete
  Appl. Math. \textbf{123} (2002), no.~1-3, 155--225.

\bibitem[BKR16]{Boothby2016}
Tomas Boothby, Andrew~D. King, and Aidan Roy, \emph{Fast clique minor
  generation in chimera qubit connectivity graphs}, Quantum Information
  Processing \textbf{15} (2016), no.~1, 495--508.

\bibitem[BPT00]{doi:10.1287/mnsc.46.7.999.12033}
Dimitris Bertsimas, Georgia Perakis, and Sridhar Tayur, \emph{A new algebraic
  geometry algorithm for integer programming}, Management Science \textbf{46}
  (2000), no.~7, 999--1008.

\bibitem[CFP01]{PhysRevA.65.012322}
Andrew~M. Childs, Edward Farhi, and John Preskill, \emph{Robustness of
  adiabatic quantum computation}, Phys. Rev. A \textbf{65} (2001), 012322.

\bibitem[Cho08]{choi}
Vicky Choi, \emph{Minor-embedding in adiabatic quantum computation: I. the
  parameter setting problem}, Quantum Information Processing \textbf{7} (2008),
  no.~5, 193--209.

\bibitem[Cho11]{Choi2011}
\bysame, \emph{Minor-embedding in adiabatic quantum computation: {II}.
  minor-universal graph design}, Quantum Information Processing \textbf{10}
  (2011), no.~3, 343--353.

\bibitem[CLO07]{Cox:2007:IVA:1204670}
David~A. Cox, John Little, and Donal O'Shea, \emph{Ideals, varieties, and
  algorithms: An introduction to computational algebraic geometry and
  commutative algebra}, Springer-Verlag, Berlin, Heidelberg, 2007.

\bibitem[CMR14]{DBLP:journals/corr/CaiMR14}
Jun Cai, William~G. Macready, and Aidan Roy, \emph{A practical heuristic for
  finding graph minors}, CoRR \textbf{abs/1406.2741} (2014).

\bibitem[CT91]{Conti:1991:BAI:646027.676734}
Pasqualina Conti and Carlo Traverso, \emph{Buchberger algorithm and integer
  programming}, Proceedings of the 9th International Symposium, on Applied
  Algebra, Algebraic Algorithms and Error-Correcting Codes (London, UK, UK),
  AAECC-9, Springer-Verlag, 1991, pp.~130--139.

\bibitem[DA17]{raouffactorization}
Raouf Dridi and Hedayat Alghassi, \emph{Prime factorization using quantum
  annealing and computational algebraic geometry}, Sci. Rep. \textbf{7} (2017).

\bibitem[DS98]{diaconis1998}
Persi Diaconis and Bernd Sturmfels, \emph{Algebraic algorithms for sampling
  from conditional distributions}, Ann. Statist. \textbf{26} (1998), no.~1,
  363--397.

\bibitem[Fau99]{Faugere199961}
Jean-Charles Faugere, \emph{A new efficient algorithm for computing {G}r\"obner
  bases ({F4})}, Journal of Pure and Applied Algebra \textbf{139} (1999),
  no.~13, 61 -- 88.

\bibitem[FGG{\etalchar{+}}01]{Farhi472}
Edward Farhi, Jeffrey Goldstone, Sam Gutmann, Joshua Lapan, Andrew Lundgren,
  and Daniel Preda, \emph{A quantum adiabatic evolution algorithm applied to
  random instances of an np-complete problem}, Science \textbf{292} (2001),
  no.~5516, 472--475.

\bibitem[JAG{\etalchar{+}}11]{natueDwave}
M.~W. Johnson, M.~H.~S. Amin, S.~Gildert, T.~Lanting, F.~Hamze, N.~Dickson,
  R.~Harris, A.~J. Berkley, J.~Johansson, P.~Bunyk, E.~M. Chapple, C.~Enderud,
  J.~P. Hilton, K.~Karimi, E.~Ladizinsky, N.~Ladizinsky, T.~Oh, I.~Perminov,
  C.~Rich, M.~C. Thom, E.~Tolkacheva, C.~J.~S. Truncik, S.~Uchaikin, J.~Wang,
  B.~Wilson, and G.~Rose, \emph{Quantum annealing with manufactured spins},
  Nature \textbf{473} (2011), no.~7346, 194--198.

\bibitem[JFS06]{PhysRevA.74.052322}
Stephen~P. Jordan, Edward Farhi, and Peter~W. Shor, \emph{Error-correcting
  codes for adiabatic quantum computation}, Phys. Rev. A \textbf{74} (2006),
  052322.

\bibitem[Kat50]{Kato}
Tosio Kato, \emph{On the adiabatic theorem of quantum mechanics}, Journal of
  the Physical Society of Japan \textbf{5} (1950), no.~6, 435--439.

\bibitem[KN98]{Nishimori}
Tadashi Kadowaki and Hidetoshi Nishimori, \emph{Quantum annealing in the
  transverse ising model}, Phys. Rev. E \textbf{58} (1998), 5355--5363.

\bibitem[Luc14]{10.3389/fphy.2014.00005}
Andrew Lucas, \emph{Ising formulations of many np problems}, Frontiers in
  Physics \textbf{2} (2014), 5.

\bibitem[MLM07]{PhysRevLett.99.070502}
Ari Mizel, Daniel~A. Lidar, and Morgan Mitchell, \emph{Simple proof of
  equivalence between adiabatic quantum computation and the circuit model},
  Phys. Rev. Lett. \textbf{99} (2007), 070502.

\bibitem[Olv99]{olver_1999}
Peter~J. Olver, \emph{Classical invariant theory}, London Mathematical Society
  Student Texts, Cambridge University Press, 1999.

\bibitem[PS01]{realalggeo}
Pablo~A. Parrilo and Bernd Sturmfels, \emph{Minimizing polynomial functions},
  DIMACS Series in Discrete Mathematics and Theoretical Computer Science
  (2001).

\bibitem[ST97]{DBLP:journals/mp/SturmfelsT97}
Bernd Sturmfels and Rekha~R. Thomas, \emph{Variation of cost functions in
  integer programming}, Math. Program. \textbf{77} (1997), 357--387.

\bibitem[Stu96]{MR1363949}
Bernd Sturmfels, \emph{Gr\"obner bases and convex polytopes}, University
  Lecture Series, vol.~8, American Mathematical Society, Providence, RI, 1996.
  \MR{1363949}

\bibitem[TTN95]{DBLP:journals/mp/TayurTN95}
Sridhar~R. Tayur, Rekha~R. Thomas, and N.~R. Natraj, \emph{An algebraic
  geometry algorithm for scheduling in presence of setups and correlated
  demands}, Math. Program. \textbf{69} (1995), 369--401.

\end{thebibliography}
}
\end{document}